\documentclass[aps,pra,10pt,twocolumn,superscriptaddress,floatfix,nofootinbib,showpacs,longbibliography]{revtex4-2}
\usepackage{makecell}
\newcolumntype{C}[1]{>{\centering\arraybackslash}m{#1}}
\usepackage{booktabs,adjustbox}
\usepackage{float}
\usepackage[normalem]{ulem}
\usepackage[utf8]{inputenc}  
\usepackage[T1]{fontenc}     
\usepackage[table]{xcolor}
\usepackage{color,soul}
\usepackage[british]{babel}  
\usepackage[sc,osf]{mathpazo}
\usepackage[scaled=0.86]{berasans}  
\usepackage[colorlinks=true, citecolor=blue, urlcolor=blue]{hyperref}  
\usepackage{graphicx} 
\usepackage[babel]{microtype}  
\usepackage{amsmath,amssymb,amsthm,bm,amsfonts,mathrsfs,bbm} 
\renewcommand{\qedsymbol}{\rule{0.7em}{0.7em}}
\usepackage{xspace}  
\usepackage{pgf,tikz}
\usepackage{xcolor}
\usepackage{multirow}
\usepackage{array}
\usepackage{bigstrut}
\usepackage{braket}
\usepackage{color}
\usepackage{natbib}
\usepackage{multirow}
\usepackage{mathtools}
\usepackage{float}
\usepackage{blkarray}
\usepackage[caption = false]{subfig}
\usepackage{xcolor,colortbl}
\usepackage{color}
\usepackage{rotating}
\usepackage{tikz}
\usepackage{tabularx}

\newcommand{\sr}[1]{\color{blue} Sap: #1}

\newcommand{\Tr}{\operatorname{Tr}}
\newcommand{\blu}{\color{blue}}
\newcommand{\grn}{\color{green}}
\newcommand{\red}{\color{red}}
\newcommand{\bla}{\color{black}}
\newcommand{\be}{\begin{equation}}
\newcommand{\ee}{\end{equation}}
\newcommand{\ba}{\begin{eqnarray}}
\newcommand{\ea}{\end{eqnarray}}
\newcommand{\ketbra}[2]{|#1\rangle \langle #2|}
\newcommand{\tr}{\operatorname{Tr}}
\newcommand{\one}{\bf{1}}
\newcommand{\half}{\frac{1}{2}}
\newcommand{\qua}{\frac{1}{4}}
\newcommand{\etal}{{\it{et. al. }}}
\newcommand{\plane}[2]{$#1#2$\nobreakdash-plane}
\newtheorem{theorem}{Theorem}
\newtheorem{corollary}{Corollary}
\newtheorem{definition}{Definition}
\newtheorem{proposition}{Proposition}

\newtheorem{protocol}{Protocol}

\newtheorem{lemma}{Lemma}

\newcommand{\N}{\mathbb{N}}
\newcommand{\Z}{\mathbb{Z}}
\newcommand{\R}{\mathbb{R}}
\newcommand{\C}{\mathbb{C}}

\newcommand{\grp}[1]{\mathsf{#1}}
\newcommand{\spc}[1]{\mathcal{#1}}

\def\d{{\rm d}}

\newcommand{\Span}{{\mathsf{Span}}}
\newcommand{\Lin}{\mathsf{Lin}}
\newcommand{\Supp}{\mathsf{Supp}}
\def\>{\rangle}
\def\<{\langle}
\def\kk{\>\!\>}
\def\bb{\<\!\<}
\newcommand{\bs}[1]{\boldsymbol{#1}}     
\newcommand{\Sp}{\mathsf{Sp}}

\newcommand{\map}[1]{\mathcal{#1}}
\newcommand{\Chan}{{\mathsf{Chan}}}
\newcommand{\Herm}{\mathsf{Herm}}
\newcommand{\Hom}{\mathsf{Hom}}
\newcommand{\End}{\mathsf{End}}
\newcommand{\Choi}{\mathsf{Choi}}

\newcommand{\St}{{\mathsf{St}}}
\newcommand{\Eff}{{\mathsf{Eff}}}
\newcommand{\Pur}{{\mathsf{Pur}}}
\newcommand{\Transf}{{\mathsf{Transf}}}
\newcommand{\Comb}{{\mathsf{Comb}}}
\newcommand{\Tester}{{\mathsf{Tester}}}
\newcommand{\Dual}{{\mathsf{Dual}}}

\newtheorem{prop}{Proposition}

\newcommand{\Proof}{{\bf Proof. \,}}
\newcommand{\tg}[1]{\textcolor{red}{#1}}
\usepackage{centernot}
\usepackage{subfig}

\begin{document}

\title{
The communication power of a  noisy qubit} 
\author{Giulio Chiribella}
\email{giulio@cs.hku.hk}
\affiliation{QICI Quantum Information and Computation Initiative, Department of Computer Science, The University of Hong Kong, Pokfulam Road, Hong Kong}
\affiliation{Department of Computer Science, University of Oxford, Parks Road, Oxford OX1 3QD,  United Kingdom}
\affiliation{Perimeter Institute for Theoretical Physics, Caroline Street, Waterloo, Ontario N2L 2Y5, Canada}
 \author{Saptarshi Roy}
 \email{sapsoy@gmail.com}
\affiliation{QICI Quantum Information and Computation Initiative, Department of Computer Science, The University of Hong Kong, Pokfulam Road, Hong Kong}
\author{Tamal Guha}
\email{g.tamal91@gmail.com}
\affiliation{QICI Quantum Information and Computation Initiative, Department of Computer Science, The University of Hong Kong, Pokfulam Road, Hong Kong}
\author{Sutapa Saha}
\email{sutapa.gate@gmail.com}
\affiliation{Department of Astrophysics and High Energy Physics, S. N. Bose National Center for Basic Sciences,
Block JD, Sector III, Salt Lake, Kolkata 700106, India}

\begin{abstract}
 A fundamental limitation of quantum communication is that a single qubit can carry at most 1 bit of classical information.  For an important class of  quantum communication channels, known as entanglement-breaking, this limitation holds even if the sender and receiver share entangled particles.
 But does this mean that,  for the purpose of communicating classical messages, 
 a noisy entanglement-breaking qubit channel can be replaced by 
 a noisy bit channel?    Here we answer the question in the negative. We introduce a game, similar to the Monty Hall problem in classical statistics, where a sender    assists a receiver in finding a valuable item (the ``prize'') hidden into one of four possible boxes,  while avoiding a hazardous item (the ``bomb'')  hidden in one of the remaining three   boxes. We show that no classical strategy using  a noisy bit channel can ensure that the bomb is avoided,  even if  the sender and receiver share arbitrary amounts of randomness. In contrast,  communication of a qubit through a class of noisy entanglement-breaking channels, which we call {\em quantum $\tt NOT$ channels},
allows the players  to deterministically avoid the bomb and to find the prize with a guaranteed nonzero probability. Our findings show that the  communication of classical messages through  a noisy entanglement-breaking qubit channel assisted by quantum entanglement  
 cannot, in general, be simulated by 
 communication through a noisy bit channel assisted by classical correlations.
 \end{abstract}
\maketitle

\textit{Introduction.}- 
The laws of quantum mechanics often give rise to unexpected phenomena that challenge our intuition and help reaching a better grasp of the subtleties of the microscopic world  \cite{Elitzur1993, Bell2014, aharonov2008quantum}.    
In this paper we present a curious phenomenon that highlights the difference between quantum and classical channels in the transmission of  classical messages:  we show that there exists  a purely classical communication task where 
the transmission of a two-dimensional quantum system through a noisy entanglement-breaking channel is more valuable than the transmission of a  two-dimensional classical system through any noisy classical  channel.

A celebrated result by Holevo \cite{Holevo1973}  implies that a $d$-dimensional quantum system, taken in isolation,  can carry at most  $\log d$ bits of classical information, where $\log$ denotes the base-2 logarithm.   This statement was later generalized by Frenkel and Weiner \cite{Frenkel2015}, who showed that the set of  conditional probability distributions achievable  by a sender and a receiver through the transmission  of a $d$-dimensional quantum system   coincides  with the set of conditional probability distributions achievable through the transmission of  a $d$-dimensional classical system, possibly assisted by correlated random variables shared by the sender and the receiver before the start of their communication.  
 This result implies that, for every possible classical task, the transmission of a $d$-dimensional quantum system can be replaced by the transmission of a $d$-dimensional classical system assisted by classical shared randomness.

The situation is different when the sender and receiver share entangled particles. In this case,  bits can be reliably transmitted through a quantum communication channel at a rate given by  the entanglement-assisted classical capacity \cite{Bennett1999}. For qubit channels, this capacity can generally exceed one bit, as famously shown by the dense coding protocol \cite{Bennett1992}.  In this case, the transmission of a two-dimensional quantum system is clearly superior to the transmission of a two-dimensional classical system.  But what if the entanglement-assisted capacity is  strictly smaller than one bit?   

\begin{figure}
    \centering
    \includegraphics[width = \linewidth]{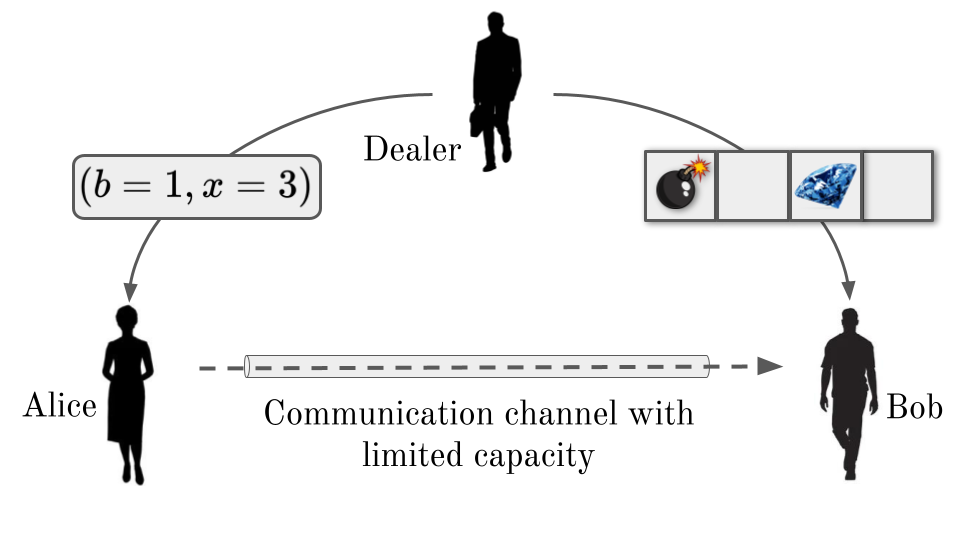}
    \caption{{\bf The bomb-and-prize game.}    
   An untrusted dealer hides a valuable item (the ``prize'') into one  of four possible boxes, labelled as $\{1,2,3,4\}$, and a hazardous item (the ``bomb'') in one of the three remaining boxes.   The dealer sends the boxes  to a player, Bob, and asks him  to open one box.   Another player,  Alice, knows the labels  $b$ and $x$ of the boxes  containing  the bomb   and the prize, respectively, and can assist Bob in finding the prize.  However,  the communication between Alice and Bob is limited to a single use of a noisy channel with limited classical capacity, even in the presence of prior correlations shared between Alice and Bob. 
    }
    \label{fig:bomb+prize}
\end{figure}
   The prototype of qubit channels with entanglement-assisted capacity  upper bounded by  1 bit is the class of {\em entanglement-breaking} channels \cite{Horodecki2003, Ruskai2003}. 
   Entanglement-breaking channels can not transmit any quantum information \cite{holevo2001evaluating, holevo2008entanglement}. 
   The transmission of a $d$-dimensional quantum system through an entanglement-breaking channel   yields at most $\log d$ bits of classical communication \cite{Fan2003}. A natural question is whether, for the purposes of classical communication,  the transmission of a two-dimensional quantum system through a noisy entanglement-breaking channel (assisted by quantum correlations) can be replaced by the transmission of a two-dimensional classical system through a noisy classical channel (assisted by classical correlations).

Here we answer the above question in the negative: we provide a purely classical  task where an entanglement-breaking qubit channel with entanglement-assisted capacity strictly smaller than $1$ bit is more valuable than  any noisy bit channel, even with  the assistance of arbitrary amounts of shared randomness between the sender and receiver. We  introduce our task in the following fictionalized scenario, which shares some similarity with the Monty Hall problem \cite{selvin1975problem} and the Three Prisoners problem  \cite{gardener1959mathematical} in classical statistics. An untrusted dealer puts a valuable item (the ``prize'') into one of four possible boxes and an hazardous item (the ``bomb")  in one of the remaining  boxes, as illustrated in Figure \ref{fig:bomb+prize}.  The dealer asks a player (Bob) to open one of the boxes. Bob's goal is to find the prize, while avoiding the bomb.  Another  player (Alice) has inside information about  the position of the bomb and prize, and can assist Bob in his decision. However, Alice and Bob are in separate locations,  and the communication between them is constrained to be strictly less than one bit. 

The primary purpose of the bomb-and-prize game  is to highlight the non-trivial relation between a peculiar features  of entanglement-breaking quantum channels and classical channels.   Nevertheless, analogues of this game may play a role  in  decision-making tasks involving limited information. For example, an investor  may have to choose between a set of possible investments, of which  one is  highly profitable and another one is a scam. Similar scenarios  could also be artificially set up  in online games, or in other recreational contexts.

Classically, the transmission of a two-dimensional system through a noisy channel is not enough to ensure that the bomb is avoided.  If Alice could communicate one bit without any noise,  then she could tell Bob whether or not the bomb is in a given subset of boxes, 
 thereby putting Bob in condition to avoid the bomb. 
But the constraint of nonunit classical capacity implies that there is  a non-zero chance that Bob opens the box containing the bomb. This conclusion does not change even if Alice and Bob share arbitrary amounts of correlated random bits: explicitly, we show that the worst-case probability of opening the box with the bomb cannot be smaller than   $(1-  C)^{\ln 4}/16$, where $C$ is the classical capacity of the bit channel connecting Alice to Bob. A similar result holds for the average probability of opening the bomb, as long as  any prior distribution that assigns non-zero probability to all possible positions of the bomb.

In stark contrast,  we show that the entanglement-assisted transmission of a qubit through a a class of noisy entanglement-breaking channels, which we call {\em quantum {\tt NOT} channels}, allows  Bob to avoid the bomb with certainty and to find the prize with a guaranteed non-zero probability. Among these channels,  the channel that maximizes the probability of finding the prize in the worst-case scenario is the  {\em universal {\tt NOT}}    ({\tt UNOT}) \cite{wernerbuzek}. This   channel allows the players to avoid the bomb with certainty and to find the prize with  probability $1/3$, despite having an entanglement assisted capacity of only $2-\log 3  \approx  0.415$ bits. The $\tt UNOT$ has been experimentally demonstrated with photons \cite{DeMartini2002,Ricci2004}, and is unitarily equivalent to the  universal transpose \cite{Gisin1999, Pawel2002}, which also has been  experimentally demonstrated \cite{Lim2011,Lim2011a,Lim2012}.  

Comparison with the classical scenario shows that simulating the noisy transmission of a qubit through any  quantum {\tt NOT} channel  requires at least a {\em noiseless}   bit channel. In fact, we find that shared randomness is also necessary:  in the absence of shared randomness, every classical strategy using a noiseless bit channel (or even a noiseless {\em trit} channel)  has   zero probability of finding the prize in the worst case scenario.  


To further highlight the role of shared randomness,  we  consider the situation  in which Alice only knows the position of the bomb. In this case, every classical strategy matching the performance of the $\tt UNOT$ strategy  through  the transmission of a single bit requires correlated classical systems of dimension at least $3$. This finding shows that there exists a classical task in which  a noisy entanglement-breaking qubit channel assisted two entangled qubits is a more powerful resource than a noiseless bit channel assisted by two correlated bits.

 We conclude by showing that a noiseless bit channel and two correlated trits  are sufficient for simulating the  {\tt{UNOT}} channel.  More generally,  every entanglement-breaking qubit channel  can be simulated by a noiseless  bit channel assisted by  shared randomness.



\medskip

\textit{Preliminaries.}- The transmission of a $d$-dimensional  quantum system from a sender to a receiver is described by a quantum channel $\mathcal{N}$, that is, a {completely positive trace preserving} map acting on $L(\mathcal{H})$, the space of all linear operators from the Hilbert space $\mathcal{H}  =  \C^d$  to itself.

   When the sender and receiver share entangled systems,  the maximum rate at which 
   they can reliably communicate classical messages   is quantified by  
   the  entanglement assisted-classical capacity \cite{Bennett1999}, given by
  $  C_{\rm E}(\mathcal{N}) := \max_{\rho} \big[ S(\rho) + S(\mathcal{N}(\rho)) -S\big(  (\mathcal{N} \otimes \map I  )  (\psi_\rho)\big)\big]$,  
where $S(\sigma) := -$Tr($\sigma \log \sigma)$ is the von Neumann entropy of a generic state $\sigma$, $\map I$ is the identity channel on $L(\spc H)$,   and the maximum is over all density matrices $\rho \in  L(\spc H)$ and over all rank-one density matrices $\psi_\rho \in L(\spc H\otimes \spc H)$  that are purifications of the density matrix $\rho$. 

For a general quantum channel on a $d$-dimensional quantum system, the entanglement-assisted capacity  can generally exceed $\log d$,  as shown by the paradigmatic example of dense coding \cite{Bennett1992}. 
However, there exists an important class of channels whose entanglement-assisted capacity is upper bounded by $\log d$.  This class consists of all  
entanglement-breaking  channels,  that is,  all quantum channels $\mathcal{N_{\rm eb}}$ such that the state $(\map N_{\rm eb}  \otimes \map I)  (  |\Psi\>\<\Psi|)$ is separable for every bipartite state $|\Psi\>$ \cite{Horodecki2003, Ruskai2003}.     For every possible entanglement-breaking channel $\map N_{\rm eb}$ acting on a $d$-dimensional quantum system,  the upper bound 
$C_{\rm E}  (\map N_{\rm eb})  \le \log d$ 
holds  \cite{Fan2003}.     If $C_{\rm E}  (\map N_{\rm eb})  < \log d$, we call  $\map N_{\rm eb}$ a {\em noisy} entanglement-breaking channel.  

\medskip

\textit{Analysis of the bomb-and-prize game.--}   The location of the bomb and prize is described by a pair $(b,x)$, where $b\in\{1,2,3,4\}$ specifies the position of the bomb and $x\in  \{1,2,3,4\}\setminus\{b\}$ specifies the position of the prize.   The overall strategy adopted by Alice and Bob can be described by a conditional probability distribution $p(y|b,x)$, where $y\in  \{1,2,3,4\}$ labels the box opened by Bob.  For a given configuration $(b,x)$, the probability of winning the game is $p(x|  b,x)$. In the following, we will consider the worst case winning probability 
\begin{align}\label{wcsucc}
p_{\rm worst}^{\rm prize}   :  =  \min_{b,x} \, p(x| b,x) \, ,
\end{align}
 under the constraint that the bomb is avoided, corresponding to the condition 
\begin{align}\label{avoidbomb0}
  p_{\rm worst}^{\rm bomb}  =  0  \quad {\rm with} \quad  p_{\rm worst}^{\rm bomb} :  = \max_{b,x}   p( b|  b,x )   \,.
\end{align}    
The choice of the worst-case scenario is motivated by the fact that the position of the bomb and prize is set by an untrusted dealer, who may act adversarially and prevent Bob from finding the prize.  In the Supplementary Material \cite{supple},  we extend our analysis to the scenario where the position of the bomb and prize is chosen at random according to a given prior distribution.   

We now show two key results about classical strategies:  {\em (i)}  the bomb-avoiding condition  (\ref{avoidbomb0}) cannot be satisfied by any noisy bit channel, and {\em (ii)}  if Alice and Bob do not share randomness, then the worst-case winning probability (\ref{wcsucc}) subject to the bomb-avoiding condition (\ref{avoidbomb0})  is zero even if a noiseless bit channel is available.  

Result {(\em i)} is established by the following theorem:  

\begin{theorem}\label{t1}
    For every classical strategy using a single-bit channel    of capacity $C  <1$ and an arbitrary  amount of shared randomness,  the probability    $ p_{\rm worst}^{\rm bomb}$ is lower bounded as
    \begin{align}
    p_{\rm worst}^{\rm bomb}  \ge  \frac{(1-  C)^{\ln 4}}{16} \, .
    \end{align}
\end{theorem}
The proof is provided in the Supplementary Material \cite{supple}, where we also extend  the result from the worst-case to  the average scenario.
  
Result {\em (ii)} is established by the following theorem:  
\begin{theorem}\label{t2}
In the absence of shared randomness between Alice and Bob, the worst-case winning probability (\ref{wcsucc}) subject to the bomb-avoiding condition (\ref{avoidbomb0}) is zero for every classical strategy using the transmission of a classical system of dimension $d\le 3$.
\end{theorem}

The  idea of the proof is to 
 show that there always exists a message $m_0$ and two  boxes $(y_1,y_2)$ that have non-zero probability to be opened if and only if Bob receives   message $m_0$. 
In the configuration where    boxes $(y_1,y_2)$ contain the bomb and  the prize, respectively,  the bomb-avoiding condition implies that Alice cannot send the message $m_0$,  and therefore Bob has no chance of finding the prize. The complete argument is presented in the Supplementary Material \cite{supple}, where we also show how this result changes when moving from the worst-case to the average scenario. 

Note that the dimension bound in Theorem \ref{t2} is tight:    if Alice could perfectly transmit a system of dimension $d=4$, then  she could communicate  the exact location of the prize  without using any shared randomness.    



\medskip  

{\em Quantum strategies.}  We now show that the transmission of a single qubit through a class of noisy entanglement-breaking channel allows  Bob to avoid the bomb with certainty and to find the prize with a guaranteed nonzero probability. We say that a qubit channel $\map N$ is a {\em {\tt NOT}  channel}  if $\map N$ is a random mixture of channels of the form $\map N_{\rm cq,  \, not} (\rho)  = |\psi_\perp\>\<\psi_\perp| \,   \<\psi|  \rho  |\psi\>  + |\psi \>\<\psi| \,   \<\psi_\perp|  \rho  |\psi_\perp\>$, where  $\{ |\psi\> , |\psi_\perp\>\}$ is an orthonormal basis.  We say that $\map N$ is a {\em quantum}  $\tt NOT$ channel if the mixture contains at least two  bases corresponding to  incompatible observables.    A prominent example of quantum {\tt NOT} channel is  the {\em (approximate) universal {\tt NOT}} \cite{wernerbuzek}, defined as 
\begin{align}\label{channel}
{\tt UNOT} (\rho):=    \frac 23 \, \Tr[\rho]  \,  I   -  \frac 13 \, \rho  \qquad \forall  \rho \in L  (\C^2) \, .
\end{align}
The $\tt UNOT$ is the quantum $\tt NOT$ channel corresponding to a uniform mixture over all possible bases, and its entanglement-assisted capacity is the lowest among all quantum $\tt NOT$ channels.    The value of the capacity can be obtained from the fact that  $\tt UNOT$ is a Pauli channel, of the form ${\map N}_{\rm pauli}  (\rho)  =  p_I  \, \rho  +  p_X  \, X\rho  X  + p_Y  \,  Y\rho  Y  + p_Z\, Z\rho Z$,  where $\{X,Y,Z\}$ are the Pauli matrices and $(p_\alpha)_{\alpha\in  \{I,X,Y,Z\}}$ is probability distribution. For Pauli channels, the entanglement-assisted capacity is  $C_{\rm E} (\map N_{\rm pauli})  = 2  +  \sum_{\alpha \in  \{I,X,Y,Z\}}   p_\alpha \log p_\alpha$ \cite{Bennett1999,bennett2002entanglement}, which in the $\tt UNOT$ case  $p_X=  p_Y=  P_Z=  1/3$ yields $2-\log 3$. 
For the reader's convenience, we include a proof of the capacity formula for Pauli channels in the Supplemental Material \cite{supple}, along with a proof that $\tt UNOT$ has minimum entanglement-assisted capacity among all $\tt NOT$ channels. 


For a quantum $\tt NOT$  channel $\map N_{\tt NOT}$,  we define the following {\em $\map N_{\tt NOT}$ dense-coding protocol}:  
\begin{protocol}[$\map N_{\tt NOT}$ dense-coding]\label{protocol:unot}
Before the start of the protocol, Alice and Bob share two qubits $A$ and $B$ in the entangled state $|\Phi^+\>_{AB}   =(|0\>_A\otimes |0\>_B  + |1\>_A\otimes |1\>_B)/\sqrt 2$. Then,  Alice applies one of the unitary gates $I,  X,  Y,$ or $Z$ depending on whether the bomb is in box $1,2,3,$ or 4, respectively.   She  sends qubit $A$ to Bob through  channel $\map N_{\tt NOT}$.  Upon receiving Alice's qubit, Bob measures both qubits $A$ and $B$ in a Bell basis, that is, the orthonormal basis consisting of states $|\Phi_1\>:  =  (I \otimes I)|\Phi^+\>,$  $|\Phi_2\>   : =  (X\otimes I) |\Phi^+\>$,  $|\Phi_3\>   : =  (Y\otimes I) |\Phi^+\>$, and  $|\Phi_4\>   : =  (Z\otimes I) |\Phi^+\>$.
 If the measurement outcome is $y$, Bob will open the $y$-th box.  
 \end{protocol}

In the Supplemental Material  \cite{supple}, we show that  the $\map N_{\tt NOT}$ dense-coding protocol has  $p_{\rm worst}^{\rm bomb}=0$  and $p_{\rm worst}^{\rm prize}>0$ for every quantum $\tt NOT$ channel $\map N_{\tt NOT}$.  In the special case $\map N_{\tt NOT} = \tt UNOT$, we find $p_{\rm worst}^{\rm prize}  =  1/3$.

More generally, one can  consider arbitrary protocols, involving arbitrary entanglement-breaking channels, entangled particles of arbitrary dimensions, arbitrary entangled states, arbitrary encoding operations, and arbitrary measurements.  A natural question is which    entanglement-breaking channels can be used to avoid the bomb  in some suitable protocol.  For qubit channels,  a complete answer is provided by the following: 
 \begin{theorem}[\cite{supple}]\label{theo:characterization}
An  entanglement-breaking qubit channel can be used to avoid the bomb in a suitable protocol if and only if it is unitarily equivalent to a $\tt NOT$ channel.  Any such channel has non-unit entanglement-assisted capacity if and only if  it is unitarily equivalent to a quantum $\tt NOT$ channel.    
 \end{theorem}
 The idea of the proof is to establish  a connection between  this game and  the tasks of quantum state exclusion 
\cite{bandyopadhyay2014conclusive, Caves2002, Pusey2012}  
and  quantum teleportation \cite{bennett1993teleporting}.  

In terms of shared entanglement, we find that  a two-qubit maximally entangled state (also known as one {\em ebit}) is  the minimal requirement for an advantage in the bomb-and-prize game:  
\begin{theorem}[\cite{supple}]\label{theo:optimality}
Every bomb-avoiding protocol using a quantum $\tt NOT$ channel requires Alice and Bob to share at least one ebit of entanglement. Every bomb-avoiding protocol using a quantum $\tt NOT$ channel and a two-qubit entangled state satisfies the bound  $p_{\rm worst}^{\rm prize}  \le 1/3$.  
\end{theorem}
 Theorem \ref{theo:optimality} establishes the optimality of the $\map N_{\tt NOT}$ dense-coding protocols in terms of shared entanglement.  Moreover, it shows that  the maximum  worst-case probability of finding the prize is $1/3$, the value achieved by the $\tt UNOT$ dense-coding protocol. Quite remarkably,   the quantum $\tt NOT$ channel with the smallest entanglement-assisted capacity offers the largest advantage in terms of worst-case winning probability. 
 See Supplementary material for details \cite{supple}.
 
\medskip  

{\em The shared randomness requirement.}    Theorem \ref{t2}  shows that, in the absence of shared randomness,  even the noiseless transmission of a trit cannot replace the transmission of a noisy qubit  in the bomb-and-prize game.  
 We now ask how much shared randomness is needed to reproduce the features of the optimal quantum  strategy, the $\tt UNOT$ dense-coding protocol. To answer this question, we introduce a modified  game in which  Alice  knows the position of the bomb, but ignores the position of the prize.  In this case,  the maximum probability of finding the prize  while avoiding the bomb is  exactly $1/3$,  and the only 
   probability  distribution $p(y|b)$ with this probability is  $p_{\tt NOT}  (y|b): =  (1- \delta_{y,b})/3$.  Hence,   matching  the performance of the $\tt UNOT$ dense-coding protocol by classical means requires simulating the $4$-dimensional classical channel $p_{\tt NOT}  (y|b)$.  In the Supplementary Material \cite{supple}, we  show that simulating $p_{\tt NOT}  (y|b)$ with a noiseless bit channel requires Alice and Bob to share correlated classical systems of dimension at least $3$:

\begin{theorem}\label{t:3}
 Every simulation of the $4$-dimensional classical  channel $p_{\tt NOT}  (y|b)$  using a bit channel and shared randomness requires the sender and receiver to share correlated classical systems of dimension $d\ge 3$. Moreover,  every simulation with $d=3$   requires  perfect correlations and maximal randomness, corresponding to $\log 3$ shared random bits.  
\label{th:th3}
\end{theorem}
A simulation protocol using   correlated trits  is provided in the Supplemental Material \cite{supple}.  


\medskip  

\textit{Simulation of entanglement-breaking qubit channels.
}  A theorem by Ruskai  shows that every entanglement-breaking qubit channel is a random mixture of  classical-to-quantum (CQ) channels, that is, channels that measure the qubit in a given basis and re-prepare it in a state depending on the measurement outcome \cite{Ruskai2003}.   
  A consequence of this fact is the following:
\begin{prop}
    Every entanglement-breaking qubit channel can be simulated through the noiseless transmission of a single bit assisted by shared randomness.
    \label{prop:1}
\end{prop}
Our results imply that the noisy version of Ruskai's theorem fails to  hold:  while  all entanglement-breaking qubit channels are random mixtures of CQ channels, there exist  entanglement-breaking qubit channels with $C_{\rm E} <1 $ that are not random mixtures of  CQ channels with  $C_{\rm E}<1 $  (see the Supplemental Material \cite{supple}).




\medskip 

\textit{Conclusions.-} We have shown that there exists a purely  classical task  in which the transmission of a qubit through  noisy entanglement-breaking channels cannot be reproduced by the transmission of a bit through a noisy bit channel, even if the sender and receiver share arbitrary amounts of randomness.  
Moreover, we have provided an example where, in the lack of shared randomness,  even  the noiseless transmission of a  three-dimensional classical system cannot reproduce the transmission of a qubit through a noisy entanglement-breaking channel.  
Our results 
can be demonstrated experimentally with photons, for example using the setups  developed in Refs. \cite{DeMartini2002,Ricci2004,Lim2011,Lim2011a,Lim2012} combined with existing techniques for the generation of Bell states and the execution of Bell measurements.

\vspace{0.5cm}
\noindent \textit{Acknowledgements.-}  GC thanks Mark Wilde for a helpful discussion on the  formula for the entanglement-assisted capacity of Pauli channels. This work has been supported by the Hong Kong Research Grant Council through grants 17307520 and R7035-21F, by the Chinese Ministry of Science and Technology through grant 2023ZD0300600, and  by the John Templeton Foundation through grant 62312, The Quantum Information Structure of Spacetime (qiss.fr).
The opinions expressed in this publication are those of the authors and do not necessarily reflect the views of the John Templeton Foundation. Research at the Perimeter Institute is supported by the Government of Canada through the Department of Innovation, Science and Economic Development Canada and by the Province of Ontario through the Ministry of Research, Innovation and Science.

\begin{widetext}

\section*{Supplementary material:
The communication power of a  noisy qubit}

\section{Proof of Theorem 1}  
\label{app:th1}
 
Let $\map B$ be the bit channel available to Alice and Bob. The bit channel $\map B$  can be represented by a conditional probability distribution  $p_{\map B}  (m'|  m)$,  $i,j\in\{0,1\}$, specifying the probability that Bob receives the message $m'  \in  \{0,1\}$ when Alice sent the message  $m\in  \{0,1\}$.   

The bit channel $\map B$ has unit capacity if and only if   $m'$ is an invertible function of $m$, that is, if and only if  $p_{\rm chan}(  m'|m)  =  \delta_{m',m}$ or $p_{\map B}(  m'|m)  =  \delta_{m',  m\oplus 1}$, where $\oplus$ denotes addition modulo 2.  Hence, the hypothesis that  $\map B$ has non-unit capacity implies that there exists at least one value $m'$ such that $p_{\map B}  (  m'  |  0)  >0$ and $p_{\map B}  (m'  |  1)  >  0$.    We call such an $m'$ an ``ambiguous  output" and denote  its  minimum probability  as 
\begin{align}\label{probambiguous}
p_{\min}   (m',  \map B)  :  = \min_m   p_{\map B}(m'|m).
\end{align}
Taking the worst-case over all ambiguous outputs, we obtain  the ambiguous probability  
\begin{align}\label{p?}
p_?  (\map B) :  =  \max_{m'}  p_{\min}   (m', \map B)   =  \max_{m'} \min_m    p_{\map B}(m'|m)   \, .
\end{align}
 Since the channel $\map B$ has non-unit capacity,  the ambiguous probability is strictly positive: $p_?   (\map B) >0$. 


Let us  start by analyzing the setting where Alice and Bob do not share randomness.   In this case, Alice's and Bob's strategy is completely described by an encoding channel $\map E$,  used by Alice to encode the position of the bomb into the input of the bit channel $\map B$, and a decoding channel $\map D$, used by Bob to convert the output of the bit channel $\map B$ into the decision as to  which box Bob should open. The two channels are represented by probability distributions $p_{\rm enc}  (m|b)$  and $p_{\rm dec}  (y|m')$,  with $b,y\in  \{1,2,3,4\}$. Hence, the probability that Bob opens box $y$ when the bomb is in box $b$ is 
\begin{align}
\nonumber p  (y|b)    &:  =  \sum_{m,m'}  p_{\rm dec}   (  y|  m')  \,  p_{\map B}(m'|m)  \,  p_{\rm enc}  ( m|  b)\\
&=   \sum_{m'}   p_{\rm dec}   (  y|  m')  \, {\sf Prob } (   m'|  b)  \, ,\label{probmn}     
\end{align}
having defined   ${\sf Prob} (   m'|  b)   :  =  \sum_{m}  p_{\map B}(m'|m)  \,  p_{\rm enc}  ( m|b)$.  

Now, the definition of $p_?  (\map B)$, provided in Eq.  (\ref{p?}),   implies that there exists a value   $m'_*$  such that $p_{\min}  (m'_*, \map B) =  p_?  (  \map B)$.  Hence, the  probability that Bob gets message $m'_*$ when  the bomb is in position $b$ is   lower bounded as 
\begin{align}
\nonumber {\sf Prob} (   m'_*|  b)    &   =  \sum_{m}  p_{\map B}(m'_*|m)  \,  p_{\rm enc}  ( m|b)    \\
  \nonumber &   \ge   \min_m     p_{\map B}(m'_*|m) \\
  \nonumber  &  \equiv  p_{\min}(  m'_* ,\map B   ) \\
      &=  p_?  (\map B) \,,
\end{align}
 independently of $b$.  

Denoting by $p_{\rm worst}:  = \max_b p (b|b)$ the  worst-case probability that Bob opens a box containing the bomb, we then have the bound  
\begin{align} 
\nonumber p_{\rm worst}    &\ge   p  (b|  b) \\
\nonumber &\ge   p_{\rm dec}   (  b|  m'_*) \, {\sf Prob }(   m'_* |  b)  \\
&\ge   p_{\rm dec}   (  b|  m'_*)   \,  p_?  (\map B)\, , \qquad \forall b\in  \{1,2,3,4\}\, . \label{utiledopo}
\end{align}  
In particular, let $y_{\max}$ be the most likely output of Bob's decoder when the output of the transmission channel is $m'_*$, so that $p_{\rm dec}  (y_{\max}  |  m'_*)  \ge  p_{\rm dec}  (y|  m'_*)$ for every $y\in \{1,2,3,4\}$.   With this definition, we have $p_{\rm dec}  (y_{\max}  |  m'_*)  \ge  1/4$, and the above bound becomes   
\begin{align}
 p_{\rm worst}    &\ge   \frac { p_?  (\map B) }  4\,    \, .  \label{boundp?}
\end{align} 
Since the non-unit capacity condition implies $p_?  (\map B)>0$,  the probability of opening the box with the bomb is nonzero.  

Bound (\ref{boundp?}) holds even  if Alice and Bob share randomness. If Alice and Bob pick  encoding and decoding operations $\map E_i$ and $\map D_i$ with probability $p(i)$, then the overall input-output distribution takes the form $p(n|m)= \sum_i  \,  p(i)\,  p^{(i)}  (y|b)$, where $p^{(i)}  (y|b)$ is the input-output  distribution for fixed encoding and decoding operations  $\map E_i$ and $\map D_i$. The worst-case probability is then lower bounded as 
\begin{align} 
\nonumber p_{\rm worst}    &\ge   \sum_i\,  p(i) \,  p^{(i)}  (b|  b) \\
\nonumber &\ge  \sum_i\,  p(i)  \,   p^{(i)}_{\rm dec}   (  b|  m'_*) \, {\sf Prob }^{(i)}(   m'_*|  b)  \\
\nonumber &\ge  \sum_i \,  p(i)\,   p_{\rm dec}^{(i)}   ( b |  m'_*)   \,  p_?  (\map B)\\
 \label{a} &  =  p_{\rm dec}  (  b |m'_*)  \,  p_?  (\map B)  \,  , \qquad \forall y\in  \{1,2,3,4\}\, ,
\end{align}  
having defined $p_{\rm dec}  (  y|m'):= \sum_i \, p(i)\,   p_{\rm dec}^{(i)}   (  y|  m')$ to be the average probability  that Bob opens box $y$ upon receiving message $m'$.  
Defining $b_{\max}$ to be the value of $b$ that maximizes $p_{\rm dec}  (  b|m'_*)$, we then have the inequality  $p_{\rm dec}  (  b_{\max}|m'_*)  \ge 1/4$, which plugged into Eq. (\ref{a}) yields Eq. (\ref{boundp?}).

To conclude, we provide a lower bound to the r.h.s. of Eq. (\ref{boundp?}) in terms of the channel capacity $C(\map B)$. To this purpose, note  that the capacity of the bit channel $\map B$  is an upper bound to  the capacity of the binary symmetric channel $ \map B_{\rm sym}$ with  probability distribution 
\begin{align}
p_{\map B_{\rm sym}}  (m'|m)  :  = \frac{ p_{\map B}   (m'|m)   +  p_{\map B}  (  m'\oplus 1  | m\oplus 1) }2 \,.
\end{align}  Explicitly, the capacity of the binary symmetric channel $\map B_{\rm sym}$ is \cite{cover1999elements}
\begin{align}
 C(\map B_{\rm sym})   =  1  -    H  (   p_{\rm sym})  \,,
\end{align}
where $H(p)  =  -p\log p  -  (1-p) \log (1-p)$ is the binary entropy, and 
\begin{align}
p_{{\rm sym}} : =       \min   \{  p_{\map B_{\rm sym}}  (0|0) ,    p_{\map B_{\rm sym}}  (1|0)     \}  \, . 
\end{align}
Since the binary entropy satisfies the bound  $H(p)  \le  [4 p  (1-p)]^{1/\ln 4} $ \cite{Topsoe2001},  we  obtain the bound  
\begin{align}
\nonumber 4  \, p_{\rm sym} \, (1-  p_{\rm sym})&\ge   [1-  C(\map B_{\rm sym})]^{\ln 4} \\
& \ge [1-  C(\map B)]^{\ln 4}  \, ,
\end{align}
which in turn implies 
\begin{align}
\nonumber p_{\rm sym}  &\ge   \frac {1  - \sqrt{ 1- [1-  C(\map B)]^{\ln 4}  }}2\\
& \ge  \frac{ [1-  C(\map B)]^{\ln 4} }{4}
\label{capacitybound} \, .    
\end{align}

We now show that  $p_{\rm sym}   \le  p_?  (\map B)$.  To simplify the notation, we define $p:  = p_{\map B}  (0|1)  $ and $q:=  p_{\map B}   (1|0)$. With this notation, we have  
\begin{align}
p_{\rm sym}  = \min\left \{   \frac{ p+q}2,  \frac{(1-p)+  (1-q)}2\right\}   \le   \min   \Big\{     \max  \{   p,  q\}   \, , \,  \max \{ 1-p,  1-q\}\Big\}  
\end{align}
and
\begin{align}
p_?   (\map B)   =  \max   \Big\{     \min  \{   p,  1-q\}   \, , \,  \min \{q  , 1-p\}\Big\}  \, .
\end{align} 
Now, there are two possible cases: either $\min \{  p, 1-q\}  =  p$, or $\min \{  p, 1-q\}  =  
1-q$.  If  $\min \{  p, 1-q\}  =  p$, then necessarily $\min \{q,  1-p\}  =  q$, and therefore 
\begin{align}
p_?  (\map B)  =   \max \{p,q\}   \ge   \min\Big\{     \max  \{   p,  q\}   \, , \,  \max \{ 1-p,  1-q\}\Big\}   \ge  p_{\rm sym} \, .
\end{align} 
If $\min \{  p, 1-q\}  =  1-q$, then necessarily $\min \{q,  1-p\}  =  1-p$, and therefore 
\begin{eqnarray}
  p_?  (\map B)  =   \max \{1-p,1-q\}   \ge   \min\Big\{     \max  \{   p,  q\}   \, , \,  \max \{ 1-p,  1-q\}\Big\}   \ge  p_{\rm sym} \, .  
  \label{eq:p?>}
\end{eqnarray}
Hence, we proved that $p_?  (\map B)  \ge  p_{\rm sym}$ in all possible cases. 
 
 To conclude the proof, we use Eqs.  (\ref{boundp?}) and (\ref{capacitybound}), which imply $p_{\rm worst}  \ge  p_?  (\map B)/4  \ge  p_{\rm sym}/4  \ge  [1-  C(\map B)]^{\ln 4} /16$. \qed

\section{Proof of Theorem 2}  
\label{app:th2}

 The most general classical strategy using a classical channel and no shared randomness consists of an encoding operation, described by the conditional probability $p_{\rm enc}(m|b,x)$ that Alice communicates the message  $m$ if the initial configuration of the boxes  is $(b,x)$,   and  a decoding operation, described by the conditional probability  $p_{\rm dec} (y|m')$ that Bob opens box $y$ upon receiving the message $m'$ from the communication channel. 
 Without loss of generality, we assume that the communication channel from Alice to Bob is the identity channel, so that $m' =  m$. Indeed, communication through any other  channel can be reproduced by appending an additional noisy operation to Bob's decoding.   
 
Now, suppose that Alice can transmit at most three classical messages, labeled by elements of the set $\{0,1,2\}$.  For every message $m\in  \{0,1,2\}$ we define the set
  ${\sf B}_m : = \{  y\, | \,  p_{\rm dec}  (y|m) \not  =  0\}$.   Since Bob opens at least one box with non-zero probability,  the set  ${\sf B}_m$ is nonempty  for every  $m \in  \{0,1,2\}$.  
  The constraint that the bomb is avoided  with certainty 
implies the condition $B_0  \cap  B_1  \cap B_2  =  \emptyset$: indeed,  if the intersection were nonempty, there would be a box that has  nonzero probability to be opened no matter what message is communicated by Alice,  meaning that there is no way to avoid the bomb with certainty if the bomb is placed into that box.  

At this point, there are two possible cases: either  (1) the union  $B_0\cup  B_1 \cup B_2$  is not the full set $\{1,2,3,4\}$, or  (2)   the union  $B_0\cup  B_1 \cup B_2$  is the full set $\{1,2,3,4\}$.  

In Case (1),   there exist a box that  has zero probability to be opened, no matter what message Alice sends.  Clearly,  putting the prize in that box brings the winning probability down to zero.  

In Case (2) there are two possible sub-cases: (2.1) there exist  two distinct values  $\alpha$ and $\beta$ such that the intersection $B_{\alpha}  \cap  B_{\beta}$ has cardinality at least $2$,  and (2.2) the intersection  $B_{\alpha}  \cap  B_{\beta}$ has cardinality at most 1 for every $\alpha\not  =  \beta $.  

In  case (2.1), we define $\gamma$ to be the remaining value, so that $\{\alpha,  \beta,  \gamma\}  =  \{0,1,2\}$.  Let $y_0$ and $y_1$ be the boxes in the intersection $B_{\alpha}  \cap B_{\beta}$.  The condition $B_0  \cap  B_1  \cap B_2  =  \emptyset$ then implies that neither $y_0$ nor $y_1$ is contained in $B_{\gamma}$.  Hence, placing the bomb in box  $y_0$ forces Alice to communicate the message $m=\gamma$ (because otherwise the box $y_0$ would have a non-zero probability of being opened), and placing the prize in box $y_1$ implies that Bob has zero probability of finding the prize  (because $y_1$ is not in  $B_{\gamma}$). 

Let us now consider the case  (2.2).   Since the intersection between any two of the sets $B_0, B_1$, and $B_2$ has cardinality at most $1$, the set $  B:  = (  B_0  \cap B_1)  \cup  (B_1  \cap   B_2  ) \cup (B_0 \cap B_2)$ has cardinality at most 3.    Hence, there exists at least one element $y_0$ that is not in $B$.    Since $B_0\cup  B_1 \cup B_2  = \{1,2,3,4\}$, the element $y_0$ should be in one of the three sets $B_0, B_1$, and $B_2$.   Let us write $y_0  \in  B_{\alpha}$ for some  value $\alpha \in  \{0,1,2\}$, and let denote by $\beta$ and $\gamma$ the two remaining values, so that    $\{\alpha,  \beta,  \gamma\}  =  \{0,1,2\}$.   At this point, there are two possible cases:   (2.2.1)  $|B_{\alpha}|  \ge 2$ and (2.2.2)  $|B_{\alpha}| = 1$.  

 In Case (2.2.1),  $B_{\alpha}$ contains at least another index $y_1$ in addition to $y_0$.  Hence,  putting the bomb in $y_1$  guarantees that Alice will not communicate the message $m=\alpha$ (because otherwise Bob would have a non-zero probability of opening the box $y_1$), and putting the prize in $y_0$ guarantees that Bob has zero probability of finding the prize (because $y_0 $ is not in $B$, and therefore it is not in the intersection   $B_{\alpha}\cap B_{\beta}$, nor in the intersection  $B_{\alpha}\cap B_{\gamma}$).

  In  Case (2.2.2),  the condition  $|B_\alpha| =1$ implies $B_\alpha  =  \{y_0\}$.  Since $y_0$ is neither in $B_\beta$ nor in $B_\gamma$  (because it is not in $B$),  the sets $B_\alpha$ and $B_\beta\cup B_\gamma$ are disjoint.  Hence,   the condition $B_\alpha\cup  B_\beta \cup B_\gamma  = \{1,2,3,4\}$ implies $B_\beta\cup B_\gamma  =  \{1,2,3,4\}\setminus\{y_0\}$.  In turn, this implies that  at least one of the  sets  $B_\beta$ and $B_\gamma$ has cardinality at least 2.  Without loss of generality, let us assume that $B_\beta$ has cardinality at least 2.  Since the intersection $B_\beta\cap B_\gamma$ has cardinality at most $1$, $B_\beta$ must contain at least one element $y_2$ that is not contained in $B_\gamma$, plus another element $y_1$ that may or may not be contained in $B_\gamma$. Putting the bomb in $y_1$ guarantees that Alice does not send the message $m=\beta$  (because otherwise Bob would have a non-zero probability of opening the box $y_1$), and putting the prize in $y_2$ guarantees that Bob has zero probability of finding the prize  (because $y_2$ is neither in $B_\alpha$ nor in  $B_\gamma$).  
  \qed

\section{From the worst-case to the average setting}

In the main text we analyzed the bomb and prize game in the worst-case scenario, motivated by the fact that the positions of the bomb and prize were selected by an untrusted party, who could also be adversarial.      Here we show how our results (Theorem 1 and Theorem 2 in the main text)   are affected if one modifies the settings from worst-case to average, that is, by assigning a prior probability distribution $\pi   (b,x)$ to the position of the bomb and prize.  We do not make any assumption on the prior, except for the basic requirement $\pi  (b,b)  =  0, \, \forall b\in  \{1,2,3,4\}$, expressing the fact that the bomb and the prize cannot be in the same box.

In the average setting, the probability to open the box with the bomb is 
\begin{align}\label{avebomb}
p^{\rm bomb}_{\rm avg}   : = \sum_{b,x}  \pi(b,x)  \,  p( b|  b,x  ) \, ,     
\end{align}
where $p(y|b,x)$ is the conditional probability distribution that Bob opens box $y$ when the bomb is in box $b$ and the prize is in box $x$. 
 
 The probability to open the box with the prize, instead, is 
\begin{align}\label{aveprize}
p^{\rm prize}_{\rm avg}   : = \sum_{b,x}  \pi(b,x)  \,  p( x|  b,x  ) \, .     
\end{align}

Theorem 1 in the main text showed that every classical strategy using a bit channel with capacity smaller than 1 will necessarily have a nonzero probability of opening the box with the bomb in the worst case.   This result remains valid also in the average scenario:   as long as the bomb as a nonzero probability to be in any of the boxes $\{1,2,3,4\}$, the probability to open the box with the bomb is nonzero even in the average scenario. A quantitative version of this statement is provided  by the following  extension of Theorem 1 in the main text: 

\begin{theorem}\label{theo:1ave}
 For every classical strategy using a single-bit channel    of capacity $C  <1$ and an arbitrary  amount of shared randomness,  the average probability    $ p_{\rm avg}^{\rm bomb}$ is lower bounded as
   \begin{align}
   \label{bombound}
   p_{\rm avg}^{\rm bomb}  \ge \pi^{\rm bomb}_{\min}  \,     \frac{[1-  C]^{\ln 4}}{4} \, ,
   \end{align}
    where $ \pi^{\rm bomb}_{\min}   :  = \min_{b\in  \{1,2,3,4\} } \,  \pi^{\rm bomb}  (b)$ and $\pi^{\rm bomb}  (b)   : = \sum_x  \,  \pi (b,x)$. 
\end{theorem}

\Proof    
Let $p_{\rm enc}  (m|b,x)$,  $p_{\rm dec}   (y|m)$, and $\map B$ be  Alice's encoding strategy, Bob's decoding strategy, and the bit channel used by Alice and Bob, respectively. 
 Then, the arguments provided in  Section \ref{app:th1} of this Supplemental Material (in particular, the derivation of Eq. (\ref{utiledopo}))  imply   that there exists a message $m^*$  such that 
\begin{align}
p(b|b,x)  \ge    p_?  (\map B)         \,      p_{\rm dec}    (b|  m_*)   \qquad \forall b\in \{1,2,3,4\},\quad \forall x\in  \{1,2,3,4\}\setminus\{b\} \, .
\end{align}
where $p_?  (\map B)$ is defined as in Eq. (\ref{p?}).  

Averaging over $b$ and $x$, we then obtain  
\begin{align}
\nonumber    p_{\rm avg}^{\rm bomb}      & \ge    p_?  (\map B)        \, \left[ \sum_{b,x}  \,  \pi  (b,x)   \,       p_{\rm dec}    (b|  m_*) \right]  \\
   \nonumber   &=    p_?  (\map B)        \, \left[ \sum_{b}  \,  \pi^{\rm bomb}  (b)   \,       p_{\rm dec}    (b|  m_*) \right]\\
     &  \ge    p_?  (\map B) \,  \pi^{\rm bomb}_{\min}        \, .
\end{align}
Combining this inequality with the bound   $p_?  (\map B) \ge  {[1-  C]^{\ln 4}}/{4}  $ (proven in Eq. \eqref{eq:p?>}  of this Supplemental Material), we finally obtain the desired result.  \qed

\medskip  

Note that the probability of opening the box with the bomb is always nonzero, except in the trivial case where the prior probability $\pi^{\rm bomb}  (b)$ is zero for some value  $b_*$.  In this case,  Alice and Bob know {\em a priori} that the bomb has zero probability to be in position $b_*$ and can avoid the bomb by opening the box in that position. 
 On the opposite end of the spectrum, it is interesting to consider the case where the probability distribution of the bomb's position is uniform, namely $\pi^{\rm bomb}  (b)  =  1/4$ for every $b\in \{1,2,3,4\}$.  In this case, the lower bound  provided by Theorem \ref{theo:1ave} in the average scenario coincides with the bound provided by Theorem 1 in the worst-case scenario.    
\medskip

We now discuss how Theorem 2 is affected if the worst-case scenario is replaced by the average one.  We consider first the case in which Alice and Bob have access to a perfect bit channel.    In this case, we have the following:  

\begin{theorem}\label{theo:2ave}
In the absence of shared randomness, the maximum of the average  winning probability (\ref{aveprize})  achievable through the communication of a $d$-dimensional classical system with $2\le d\le 4$,   subject to the bomb-avoiding condition $p_{\rm avg}^{\rm bomb}  = 0$, is equal to 
\begin{align}\label{maxaveprize}
p_{\rm avg,  \max}^{\rm bomb} =   \sum_{m=1}^{d-1}  \,  \pi^{\rm prize \, \downarrow}  (m)  \, ,
\end{align}
where $\pi^{\rm prize}  (x)   : =  \sum_b  \,   \pi  (b,x)$ is the probability distribution of the prize, and $\pi^{\rm prize \, \downarrow}   (m)$ is the $m$-th largest entry of $\pi^{\rm prize}  (x)$.   
\end{theorem}
\Proof  Since the probabilities  $p_{\rm avg}^{\rm prize}$ and  $p_{\rm avg}^{\rm bomb}$ are  linear in   the encoding and decoding channels,  the maximization can be restricted without loss of generality to deterministic encoding and deterministic encoding, specified by two functions $m=  f_{\rm enc}  (b,x)$ and $y=  f_{\rm dec}  (m)$.   

 Upon receiving message $m$,  Bob will open box $y_m   :=  f_{\rm dec}  (m)$.   Hence, the probability that Bob finds the prize is 
 \begin{align}
\nonumber p^{\rm prize}_{\rm avg}   &: = \sum_{m} \,  \sum_{b:   f_{\rm enc}  (b,  y_m) =m  }  \pi(b,y_m)\\
\nonumber  &  \le   \sum_m    \sum_{b\not  =  y_m }  \pi(b,y_m)  \\
 &   = \sum_m \, \pi^{\rm prize}  (y_m)    \, .       \label{sumtwoprob}
\end{align}
The equality sign  can be attained by an  encoding function that satisfies the bomb avoiding condition.  For example, one such encoding can be constructed by assigning the value $m$ to all the pairs $(b,x)$ in the set $A_m$, defined as
\begin{align}
\nonumber A_m    &:=    \Big\{    (b,  y_m) ~|~  \forall b\not  =  y_m \Big  \}  \cup  \Big\{  (y_{m + 1},x)~|~ \forall  x  \not  \in   \{y_1,\dots,  y_{d}\}   \Big\}      \qquad \forall m  \le d-1 \\ 
A_d &:  =    \Big\{    (b,  y_d) ~|~  \forall b\not  =  y_d \Big  \}   \cup  \Big\{  (y_1,x)~|~ \forall  x  \not  \in   \{y_1,\dots,  y_{d}\}   \Big\}      \cup  \Big\{  (b,x)~|~ \forall  b, x  \not  \in   \{y_1,\dots,  y_{d}\} \, , \forall b\not  =  x  \Big\}\, .
\end{align}
By maximising the r.h.s. of Eq. (\ref{sumtwoprob}) over the choice of $y_1$,  $y_2,  \dots,  y_d$ we then obtain Eq. (\ref{maxaveprize}). \qed

\medskip  
Theorem \ref{theo:2ave} shows a different behaviour compared to the worst-case scenario:  while the worst-case winning probability is zero for every $d\le 3$,  the average winning probability is strictly positive for every $d\ge 2$.   From Eq. (\ref{sumtwoprob})  we can see that   the least favourable prior to Alice and Bob  is the uniform prior $\pi^{\rm prize}   (x) = 1/4  \forall   x\in  \{1,2,3,4\}$: in this case, the success probability is $1/2$ for $d=2$, and $3/4$ for $d=3$  (for $d=4$ the success probability is 1, but this also happens in the worst-case scenario because Alice and just communicate to Bob the position of the prize).     Hence, in the average scenario  perfect classical communication of a bit or a trit is more beneficial that access to the universal {\tt NOT} channel, which only yields a 1/3 winning probability.  Note also that shared randomness is useless in the average scenario, because the linearity of the figure of merit implies that the optimization can be restricted to pure strategies.    

From a foundational point of view, the worst-case scenario appears to be a more powerful lens for highlighting the potential of noisy quantum communication:  this setting provides an instance of a task where the noisy transmission of a qubit is more powerful than the noiseless transmission of a trit, if no shared randomness is available. 
 This feature is somehow ``washed away'' by the average over a prior, which makes shared randomness unnecessary and allows noiseless classical channels to beat the noisy quantum channel used in our protocol.   Still, a neat quantum advantage remains even in the average setting:  while a noisy qubit channel assisted by quantum entanglement can ensure that the bomb with certainty,  every noisy bit channel assisted by shared randomness will necessarily lead to a non-zero probability of opening the box with the bomb, as shown by Eq.     (\ref{bombound}).

\section{Basic facts about entanglement-assisted capacities}

\subsection{Entanglement-assited capacity of Pauli channels and other covariant channels}
The explicit expression for the entanglement-assisted capacity of Pauli channels and other covariant channels  can be derived from an argument provided in the original papers \cite{Bennett1999,bennett2002entanglement}. This expression  has often appeared in the literature \cite{liang2002entanglement, pirandola2017fundamental, chandra2022entanglement}.  
For the convenience of the reader, we reproduce a proof here.

We will provide the argument in its general form, applying to quantum channels that are covariant with respect to the action of a group, in the following sense:  
\begin{definition}
 Let $\grp G$ be a  group and let $\spc H$ be a Hilbert space.  A function $U:  \grp G \to  L(H_{A})  , \,   g  \mapsto  U_g $ is a projective  unitary representation of $\grp G$ if   
 \begin{enumerate}
 \item for every $g\in \grp G$, $U_g$ is a unitary operator  
 \item  $U_e=  I$ for the identity element $e\in  \grp G$
 \item for every $g,h\in \grp G$, $U_g U_h  =  \omega(g,h) \,  U_{gh}$, where  $\omega(g,h)$ is a complex number with $|\omega (g,h)|=1$.   
 \end{enumerate}
 \end{definition}

 \begin{definition}
  Let $\grp G$ be a group, and let $U:  \grp G \to  L(H_{A})  , \, g  \mapsto  U_g $  and $V:  \grp G \to  L(H_{B})  , \, g  \mapsto  V_g$  be two projective unitary representations of $\grp G$ on two finite-dimensional Hilbert spaces  $\spc H_A$ and $\spc H_B$, respectively.  A quantum channel $\map N:  L(\spc H_A) \to L(\spc H_{B})$ is {\em $(U,   V)$-covariant} if 
 \begin{align}
\map N\circ \map U_g  =  \map V_g \circ \map N \qquad \forall g\in \grp G \, ,
\end{align}
where   $\map U_g$ and $\map V_g$ are the unitary channels associated to the operators $U_g$ and $V_g$, respectively. 
 \end{definition}
 
In general, the entanglement-assisted capacity of an arbitrary channel $\map N:  L(\spc H_A) \to L(\spc H_B)$ is given by \cite{Bennett1999}
\begin{eqnarray}
    C_{\rm E}  (\map N)     =  \max_{  \rho \in L(  \spc H_A)   , \, \rho\ge 0  ,\, \Tr[\rho]=1} I(\rho, \map N),
\end{eqnarray}
where $I(\rho, \map N) = S  ( A:B  )_{(\map N\otimes \map I_A)  (|\Psi_\rho\>\<\Psi_\rho|)}$,  $|\Psi_\rho\>\<\Psi_\rho|  \in L  (\spc H_A \otimes \spc H_A)$ is a purification of $\rho$, and for a generic state $\sigma_{AB} \in  L(\spc H_{A} \otimes \spc H_B)$, $S(A:B)_{\sigma_{AB}}$ is the quantum mutual information 
\begin{align}
S(A:B)_{\sigma_{AB}} :  =  S (\sigma_A)  + S  (\sigma_B)  -  S  (\sigma_{AB}) \, ,
\end{align}
with   $\sigma_A: = \Tr_B[\sigma]$, $\sigma_B: =  \Tr_A[\sigma]$, and     
$S  (\sigma_X)  : =  -\Tr[\sigma_X \log \sigma_X]$ for $X  \in  \{ A,  B,  AB \}$.   

When a quantum channel is covariant, it is easy to see that the maximization of the mutual information can be restricted without loss of generality to purifications of quantum states that are invariant with respect to the group action, that is, states  $\rho \in L(\spc H_A)$ satisfying the invariance condition 
\begin{align}
U_g \rho  U_g^\dag=  \rho  \, ,\qquad \forall g\in\grp G\, ,
\end{align}
or equivalently, the commutation relation 
\begin{align}
[\rho,  U_g]  =  0  \, , \qquad \forall g\in \grp G \, .
\end{align}
 \begin{lemma}\label{lem:symmetric}
 Let $\grp G$ be a finite or compact group,  and let $U:  \grp G \to  L(H_{A})  , \, g  \mapsto  U_g $  and $V:  \grp G \to  L(H_{B})  , \, g  \mapsto  V_g $  be two projective unitary representations of $\grp G$ on two finite-dimensional Hilbert spaces  $\spc H_A$ and $\spc H_B$, respectively.  If a channel $\map N:  L(\spc H_A) \to L(\spc H_B)$ is $(U,V)$-covariant, 
then 
\begin{eqnarray}\label{symmetric}
    C_{\rm E}  (\map N)     =  \max_{  \rho \in L(  \spc H_A)   ,\, \rho\ge 0  ,\, \Tr[\rho]=1 , \, [\rho,  U_g] = 0  \, \forall g\in  \grp G } I(\rho, \map N),
\end{eqnarray}
\end{lemma} 
\Proof  For a compact group $\grp G$, we denote by  $\int \d g $  the integral over the normalized Haar measure, with the understanding that the integral is a sum in the case of finite groups.    

For a generic  quantum state $\rho  \in  L(\spc H_A)$, the state 
\begin{align}
\<\rho\>  :=  \int \d g  \,  U_g \rho  U_g^\dag  
\end{align}
satisfies the invariance condition $[\rho,  U_g]  =   0 \, , \forall g\in \grp G$.  

Now, the mutual information  $I(\rho, \map N)$  is concave in  $\rho$  \cite{Wilde2011}.  Hence, we have 
\begin{align}
\nonumber I(\, \<\rho\>,  \,  \map N ) & \ge  \int \d g \,  I(  U_g \rho  U_g^\dag  ,  \map N )  \, .         
\end{align}
On the other hand, every purification of the state $U_g\rho  U_g^\dag$ is of the form $(U_g  \otimes I)  |\Psi_\rho\>$, where $|\Psi_\rho\>$ is a purification of $\rho$. Hence, we have the equality  
\begin{align}
\nonumber I(  U_g \rho  U_g^\dag  ,  \map N ) &  =     I(   \rho    ,  \map N\circ \map U_g )\\ 
\nonumber &    =  I(\rho,  \map V_g\circ \map N)  \\
&  =  I(\rho,   \map N)  \qquad \forall g\in\grp G \, ,
\end{align}
where the second equality follows from the covariance of the channel, and the third equality follows from the invariance of the quantum mutual information under unitary channels acting on system $B$.  

Summarizing, we obtained the inequality 
\begin{align}
 I(\rho,  \map N) \le I( \, \<\rho  \>  \,  ,  \map N)      \qquad \forall \rho  \in  L(\spc H_A)\,,
\end{align}
which implies Eq.  (\ref{symmetric}). \qed  

\medskip  

 Lemma \ref{lem:symmetric} yields a simple formula when the representation $(U_g)_{g\in\grp G}$ is irreducible, that is, when the condition $[\rho,  U_g]  =  0$ implies $\rho = I_A/d_A$ for every quantum state $\rho \in  L(\spc H_A)$.  In this case, the maximum  of the mutual information is achieved by a purification of the maximally mixed state.   

\begin{proposition}\label{prop:irreducible}
 Let $\grp G$ be a finite or compact group, let $U:  \grp G \to  L(H_{A})  , \, g  \mapsto  U_g $  and $V:  \grp G \to  L(H_{A})  , \, g   \mapsto  V_g $  be two projective unitary representations of $\grp G$ on two finite-dimensional Hilbert spaces $\spc H_A$ and $\spc H_B$, respectively, and let $\map U_g$ and $\map V_g$ be the unitary channels associated to the operators $U_g$ and $V_g$, respectively.  If a quantum channel $\map N:  L(\spc H_A) \to L(\spc H_{B})$ is $(\map U,  \map V)$-covariant and the representation $\map U$ is irreducible, then
 \begin{align}
 C_{\rm E}   (\map N)   =  S (A:  B)_{\Choi (\map N)} \, , 
 \end{align}
 where $\Choi  (\map N)$ is the Choi state \cite{Choi1975} 
 \begin{align}\label{cechoi}
 \Choi (\map N)  :  =  ( \map I_A \otimes \map N )  (|\Phi^+\>\<\Phi^+|) \, , 
 \end{align}
 $|\Phi^+\>:  = \sum_{n=0}^{d_A-1}  \,  |n\> \otimes |n\>/\sqrt {d_A}$ being the canonical maximally entangled state in $\spc H_A\otimes \spc H_A$. 
\end{proposition}

\Proof Immediate from Lemma \ref{lem:symmetric} and from the fact that an irreducible representation admits only one invariant state, namely the maximally mixed state $\rho  =  I_A/d_A$.   \qed  

\medskip  

Pauli channels are a special case of   channels that are covariant with respect to an irreducible representation on the input system.  In this case, the group is the Abelian group  $\Z_2 \times \Z_2$, with the projective representation  $U:  (p,q) \in \Z_2\times \Z_2 \mapsto  U_{pq} \in L(\C^2)$ defined by 
\begin{align}
U_{00}  :=  I  , \quad U_{01}  :=  Z  , \quad  U_{10}  : =  X  ,  \quad U_{11}  :=   XZ  \equiv -i Y \, .  
\end{align}
More generally, one can consider the group $\Z_d \times \Z_d$ and the projective representation   $U:    (p,q) \in \Z_d \times Z_d \mapsto U_{p,q}  \in  L  (\C^d)$ defined by 
\begin{align}
U_{pq}  :=    S^p  M^q \, ,   \qquad \forall p,q \in  \{0,1,\dots, d-1\} \, ,
\end{align}
where 
\begin{align} 
S  = \sum_{n=0}^{d-1}  \,  |(n+1) \mod d \>\< n|  \qquad{\rm and}  \qquad  M  =  \sum_{n=0}^{d-1}  \,  e^{\frac{  2\pi  i   n }d} \,  |n\>\<n|   
\end{align}
are the shift and multiply operators, respectively.  

The discrete Weyl-Heisenberg representation is irreducible for every $d\ge 2$.  Examples of Weyl-Heisenberg covariant channels were provided in \cite{chiribella2005extremal}. 
An interesting subset of Weyl-Heisenberg covariant channels, containing all Pauli channels in the $d=2$ case, is the set of Bell-diagonal channels \cite{bennett2002entanglement}:  
\begin{definition}
A {\em Bell-diagonal channel} is a random unitary channel $\map N_{\vec \lambda}$ of the form 
\begin{align}\label{nbell}
{\map N_{\vec \lambda}}  (\rho)  : =  \sum_{p,q\in\{0,\dots, d-1\}}
 \,  \lambda_{ pq} \,     U_{pq} \,  \rho  \,  U_{pq}^\dag    \qquad \forall \rho \in  L(\C^d)\, ,
 \end{align}
 where $\vec \lambda : =  (\lambda_{pq})_{p,q\in  \{0,\dots, d-1\}}$ is a probability distribution.    
 \end{definition}  
In the $d=2$ case, the above definition coincides with the definition of the Pauli channels.   

For every Bell-diagonal channel $\map N_{\vec \lambda}$, the Choi state is diagonal in the Bell basis, namely  
\begin{align}
 \Choi  (\map N_{\vec \lambda})   =  \sum_{p,q}  \, \lambda_{pq}  \,  |\Phi_{pq}\>\<\Phi_{pq} |  \, ,  
\end{align}
where  $|\Phi_{pq}\>   :  =  (I \otimes U_{pq})  |\Phi^+\>$,  $p,q \in  \{0,\dots, d-1\}$ are the canonical Bell states. 
Since the marginals of the Bell states are maximally mixed, the mutual information of the Choi state is 
\begin{align}
S(A:B)_{\Choi  (\map N_{\vec \lambda}) }     =   2\log d  -  \sum_{p,q}  \, \lambda_{pq}  \log \lambda_{pq} =: H(\vec \lambda)\, .     \end{align}

Hence, we obtained the following 
\begin{corollary}\label{prop:ceabell}
For a general Bell-diagonal channel $
\map N_{\vec \lambda}$, the entanglement-assisted capacity   is 
\begin{align}\label{ceabell}
C_{\rm E}  (\map N_{{\vec \lambda}})   = 2 \log d  -  H  (\vec \lambda)   \,   \end{align}
where $H(\vec \lambda)$ is the Shannon entropy  of the probability distribution $\vec \lambda$. 
\end{corollary}  

In the special case of a Pauli channel $\map N_{\rm pauli}  (\rho)  =  p_I  \, \rho  +  p_X  \rho X  +  p_Y  \,  Y\rho Y  +  p_Z \,  Z\rho  Z$, Eq. (\ref{ceabell}) becomes 
\begin{align}\label{ceapauli}
C_{\rm E}  (\map N_{\rm pauli})  =  2   +  \sum_{\alpha \in  \{  I,X,Y,Z\} }  p_\alpha \log p_\alpha \, .
\end{align}

\subsection{Entanglement-assisted capacity of the $\tt UNOT$ channel}
In particular, the $\tt UNOT$ channel 
 is the Pauli channel with $p_X  = p_Y  =  p_Z   =1/3$, and Eq. (\ref{ceapauli}) yields 
 \begin{align}
 C_{\rm E}  ({\tt UNOT})  =     2 -  \log 3  \, .    
 \end{align}

Later in this Supplemental Material we will show that the  entanglement-assisted capacity of the $\tt UNOT$ channel is minimal among all the $\tt NOT$ channels, that is, all convex combinations of  channels $\map N_{\rm cq, \, not}$ of the form  
\begin{align}\label{ncqnot}
\map N_{ \rm cq, not } (\rho)   =  |\psi_\perp\>\<\psi_\perp|  \,  \<\psi| \rho |\psi\>   + |\psi \>\<\psi|  \,  \<\psi_\perp| \rho |\psi_\perp\>    \,,  \qquad \forall \rho
\end{align}
where $\{|\psi\>,|\psi_\perp\>\}$ is an orthonormal basis.  

The argument uses the following lemma: 
\begin{lemma}\label{lem:avenot}
For a qubit channel $\map N$, let $\<\map N\>$ be the qubit channel defined by 
\begin{align}
\<  \map N\>   :  =  \int  \d U  \, ~  \Big(  \map U  \circ  \map N  \circ \map U^\dag\Big) \, ,
\end{align} 
where  $U:  \C^2 \to \C^2$ is a unitary operator,  $\d U$ is the normalized Haar measure over the unitary group, $\map U$ is the unitary channel associated to $U$, and $\map U^\dag=  \map U^{-1}$ is the unitary channel associated to $U^\dag$.  
Then,  for every $\tt NOT$   channel $\map N_{\tt NOT}$,   one has
\begin{align}
\<  \map N_{\tt NOT}  \>      =  \tt UNOT \, . 
\end{align}
\end{lemma}
\Proof  Since every $\tt NOT$ channel is a mixture of channels $\map N_{\rm cq, \, not}$ of the form (\ref{ncqnot}),  and since the map $\map N  \mapsto \<\map N\>$ is linear, it is enough to prove the thesis for a single channel  $\map N_{\rm cq, \, not}$.    For such a channel, we have 
\begin{align}
\<  \map N_{\rm cq, \, not}  \>   =  2\int \d \psi  ~   |\psi_\perp\>\<\psi_\perp|    \,   \Tr \big[  \rho  \,  |\psi\>\<\psi|  \big] \,,   \label{integralunot}   
\end{align}
where $\d \psi$ is the normalized invariant measure on the set of all unit vectors.   On the other hand, Ref. \cite{chiribella2021quantum}
 showed that the integral (\ref{integralunot}) is equal to ${\tt UNOT}  (\rho)$.  Since $
 \rho$ is arbitrary, we conclude $\<\map N_{\rm cq,  \,not}\>  =  \tt UNOT$ for every channel of the form (\ref{ncqnot}), and, by linearity $\<  \map N\>  =  \tt UNOT$ for every mixture of such channels, that is, for every $\tt NOT$ channel.  \qed

 \medskip  

 Using the above lemma, it is easy to show that the $\tt UNOT$ channel has the smallest entanglement-assisted capacity among all $\tt NOT$ channels: 
 \begin{proposition}
 For every $\tt NOT$ channel $
\map N_{\tt NOT}$, one has the bound 
\begin{align}
 C_{\rm E}  (\map N_{\tt NOT})  \ge  C_{\rm E}    \Big(\, {\tt UNOT}\,  \Big) \, . 
\end{align}
\Proof The bound follows from the following inequalities: 
\begin{align}
\nonumber  C_{\rm E}    ({\tt UNOT})   &  =  C_{\rm E}   (  \<  \map N_{\tt NOT}\>)  \\
\nonumber  
&  = C_{\rm E}  \left(   \int \d U\, ~ \map U\circ \map N_{\tt NOT} \circ  \map U^\dag \right)\\
\nonumber &  \le \int \d U\, ~  C_{\rm E}  \left(    \map U\circ \map N_{\tt NOT}  \circ \map U^\dag \right)\\
&=  C_{\rm E}  (\map N_{\tt NOT} )\, ,
\end{align}
where the first equality follows from Lemma \ref{lem:avenot}, the first inequality follows from the convexity of the entanglement-assisted capacity as a function of the channel \cite{Wilde2011}, and the last equality follows from the condition $C_{\rm E}  \left(    \map U\circ \map N_{\tt NOT}  \circ \map U^\dag \right)   =  C_{\rm E}  \left(  \map N_{\tt NOT} \right)\, , \forall U$.   
 \end{proposition}

\section{Analysis  of the  $\map N$-dense-coding protocol }

Here we show that the $\map N$-dense coding-protocol described in the main text allows Bob to avoid the bomb with certainty and to win the prize with a guaranteed nonzero  probability in the worst-case scenario.

Let us start from the case of the $\tt UNOT$ channel. 
Suppose that the bomb is in position $b\in  \{1,2,3,4\}$.  Then, Alice's encoding operation will take qubits $A$ to the state  $|\Phi_b\>$. After qubit $A$ has been sent through the {\tt UNOT} channel, the final state of  qubits $AB$ is a uniform mixture of the three Bell states orthogonal to  $|\Phi_b\>$, namely
\begin{align}
\nonumber ({\tt UNOT} \otimes  \map I_B )  (  |\Phi_b\>\<\Phi_b|)   &=  \frac 13  \,  I  \otimes I    -  \frac 13  \,    |\Phi_b\>\<\Phi_b| \\
&  = \frac 13  \,\sum_{y\not  =  b} |\Phi_y\>\<\Phi_y|\, . 
\end{align}
Hence, when  Bob measures the two qubits in the Bell basis, he will obtain an outcome $y$ that is guaranteed to satisfy the condition   $y\not = b$.   Thanks to this fact, Bob avoids the bomb with certainty.  Moreover, since the value of $y$ is uniformly random in the set $\{1,2,3,4\}\setminus \{b\}$, Bob is guaranteed to find the prize with probability $p_{\rm worst}^{\rm prize}  =  1/3$.     

The net effect of the {\tt UNOT} dense coding protocol is to reproduce the transmission of a four-dimensional classical system through the  {\em  $4$-dimensional {\tt NOT} channel},  specified by the conditional probability distribution 
\begin{align}\label{classicalNOT}
p_{\tt NOT}   (y|b)  =  \frac 13\,     (1-  \delta_{y,b}) \, ,  \qquad\forall y,b
\end{align}
where $\delta_{y,b}$ is the Kronecker delta.

It is interesting to observe that the $4$-dimensional {\tt NOT} channel  has capacity    $C =  2-\log_2 3$-bits, exactly equal to the entanglement-assisted capacity of the universal $\tt NOT$ channel.    On the other hand, Theorem 1 in the main text proved that the transmission of a two-dimensional classical system through a noisy channel of capacity $C<1$ is not sufficient to avoid the bomb with certainty.    Since the classical  $4$-dimensional {\tt NOT} channel  does guarantee a zero probability to open the box with the bomb, we conclude that this $4$-dimensional channel cannot be simulated by any  noisy  2-dimensional channel, even though its communication capacity is strictly smaller than one bit.   

 Let us consider now an arbitrary quantum $\tt NOT$ channel $\map N$.   Later in this Supplemental Material we will show that every quantum $\tt NOT$ channel is {\em equivalent to the $\tt UNOT$ channel under stochastic degradation}, meaning that there exist two non-zero probabilities $\lambda>0$ and $\mu>0$ and two qubit channels $\map C$ and $\map D$ such that  
\begin{align}
{\tt UNOT   }  =  \lambda\,  \map N  +   (1-\lambda) \, \map C \qquad {\rm and} \qquad   \map N  = \mu \,  {\tt UNOT} +  (1-\mu)\,  \map D   
\end{align}
(see Proposition \ref{prop:quantumnotchannels} for the proof).  

 Now, for a generic qubit channel $\map M$, let us denote by $p^{\rm dc}_{\map M}  (y|b,x)$  the probability of opening box $y$ in the $\map M$-dense-coding protocol when the bomb is in box $b$ and the prize is in box $x$. Then, we have the relations
  \begin{align}
\nonumber    0   &=  p^{\rm dc}_{\tt UNOT}   (b| b,x  ) \\
\nonumber &= \lambda\,   
p^{\rm dc}_{\map N} ( b|b,x) +  (1-\lambda)\,  p^{\rm dc}_{\map C } ( b|b,x) \\
 \nonumber &   \ge  \lambda   p^{\rm dc}_{\map N} ( b|b,x) \, ,  \qquad \forall b,x
\end{align}
 and
 \begin{align}
\nonumber p^{\rm dc}_{\map N}   (x| b,x  )     &= \mu\,   p^{\rm dc}_{\tt UNOT} ( x|b,x) +  (1-\mu)\,  p^{\rm dc}_{\map D} ( x|b,x) \\
 \nonumber &   \ge  \frac \mu 3 \qquad \forall b,x
\end{align}
which imply $p_{\rm worst}^{\rm bomb}  = 0$ and $p_{\rm worst}^{\rm prize}  >  0$ for the $\map N$-dense-coding protocol, respectively.

\section{Proof of  Theorem 3}

The proof proceeds by  considering all possible quantum strategies for the bomb-and-prize game, and by  characterizing the strategies that satisfy the bomb-avoiding condition.  
\begin{definition}
A {\em quantum bomb-and-prize strategy}  consists of 
\begin{itemize}
\item the transmission of a quantum system $Q$ from Alice to Bob through a  channel $\map N$ 
\item  quantum systems $A$ and $B$ of arbitrary dimensions, located in  Alice's and Bob's laboratories, respectively, and initialized in an arbitrary  bipartite state $\rho_{AB}$ 
\item arbitrary encoding operations $\map E_{b,x}   :  L(  \spc H_A) \to L(\spc H_Q) $, used by Alice to encode the positions of the bomb and prize into the input of channel $\map N_{\rm eb}$,   
\item arbitrary joint  measurements on system $QB$, corresponding to POVMs $(  M_y )_{y\in  \{0,1,2,3\}}$, $M_y  \in L  (\spc H_Q \otimes \spc H_B)$, used by Bob to decide which box $y$ he  will  open.
\end{itemize}
\end{definition}

In the following, a quantum bomb-and-prize strategy  will be represented by the quadruple $  ( \map N ,    \rho_{AB}  ,  (\map E_{b,x})_{b,x}    \, ,  (M_y)_{y}  )$.  
For a given state, encoding operation, and measurement,  the probability that Bob opens box $y$ when the bomb is in $b$ and the prize is in $x$ is a linear function of the channel $\map N$, given by 
\begin{align}\label{pybxquantum}
p_{\map N}( y|b,x)   :  = \Tr  [  M_b   \,    (\map N \map E_{b,x}  \otimes \map I_B)  (\rho_{AB}) ]  \, .
\end{align}
Hence, the bomb-avoiding condition reads 
\begin{align}\label{quantumbombavoid}
p_{\map N}   (b|b,x)  \equiv  \Tr  [  M_b   \,    (\map N \map E_{b,x}  \otimes \map I_B)  (\rho_{AB}) ]   =  0  \qquad \forall b\in  \{1,2,3,4\}  \, , \forall x\in  \{1,2,3,4\}\, .
\end{align}
\begin{definition}
 We say that a quantum channel $\map N: L(\spc H_Q) \to L(\spc H_Q)$ {\em satisfies the bomb-avoiding condition} if Eq.  (\ref{quantumbombavoid}) holds  for some  quantum systems $A$ and $B$, some bipartite state $\rho_{AB}$, some set of encoding operations $\{\map E_{b,x}\}_{b, x}$, transforming system $A$ into system $Q$,  and some POVM $(  M_y)_y$, representing a joint measurement on systems $QB$.     
\end{definition}

In the following, we will focus our attention to strategies where the channel $\map N$ is entanglement-breaking.   Recall that every entanglement-breaking channel $\map N_{\rm eb}$ is of the measure-and-prepare form \cite{Horodecki2003}  
\begin{align}\label{measprep}
\map N_{\rm eb}  (\rho)   =  \sum_{i   =  1}^N  \,  \rho_i  \,  \Tr [  \rho  P_i] \, ,      
\end{align}
where $N$ is a positive integer, $(P_i)_{i=1}^N$ is a POVM,
 and  $\rho_i$ is a normalized density matrix for every $i\in  \{1,\dots, N\}$. 

The following lemma establishes a connection between bomb-and-prize strategies and teleportation-like schemes:
 \begin{lemma}\label{lem:telechan} 
 An entanglement-breaking channel  (\ref{measprep})   satisfies the bomb-avoiding condition  (\ref{quantumbombavoid}) if and only if  
\begin{align}\label{exclusion}
  \Tr \big[ P_i  \,  \map T_x  (\rho_i)\big]=  0  \qquad \forall i\in  \{1,\dots,  N\} \, ,\forall x\in \{1,2,3,4\}  \, ,
\end{align}
where $\map T_x:  L(  \spc H_Q )\to L(\spc H_Q)$ is the quantum channel defined by 
\begin{align}
\map T_x  (\rho)   :=  \sum_b  \,    \map E_{b,x} \left(  \Tr_{QB}  \Big[(  M_b\otimes I_A)   \, ( \rho \otimes \rho_{AB} )  \Big] \,   \right)  
\end{align}
(in the r.h.s. it is understood that the Hilbert spaces are suitably ordered according to their labels.)
\end{lemma}
\Proof For an entanglement-breaking channel as in Eq. (\ref{measprep}), the bomb avoiding condition can be rewritten as 
\begin{align}
\sum_{i=1}^N \,  \Tr  \left\{  M_b   \,    \left(\rho_i \otimes
 \Tr_Q  \left[   (P_i \otimes I_B  ) \, (\map E_{b,x}  \otimes \map I_B)  (\rho_{AB}) \right]\right) \,\right\}  =  0  \qquad \forall b\in  \{1,2,3,4\}  \, , \forall x\in  \{1,2,3,4\}\, ,
\end{align}
or equivalently, 
\begin{align}
\sum_{i=1}^N \,  \Tr \left\{  P_i  \, \map E_{b,x}\left(    \Tr_{QB}    \left[  (M_b \otimes I_A) (\rho_i \otimes \rho_{AB}) \right] \right)\right\}  =  0  \qquad \forall b\in  \{1,2,3,4\}  \, , \forall x\in  \{1,2,3,4\}\, .
\end{align}
Since all the terms in the sum are nonnegative, the above condition is equivalent to
\begin{align}
\Tr \left\{  P_i  \, \map E_{b,x}\left(    \Tr_{QB}    \left[  (M_b \otimes I_A) (\rho_i \otimes \rho_{AB}) \right] \right)\right\}  =  0  \qquad \forall b\in  \{1,2,3,4\}  \, , \forall x\in  \{1,2,3,4\}\, \forall i\in \{1,\dots,  N\} \, ,
\end{align}
which in turn is equivalent to the equation 
\begin{align}
\sum_b  \,  \Tr \left\{  P_i  \, \map E_{b,x}\left(    \Tr_{QB}    \left[  (M_b \otimes I_A) (\rho_i \otimes \rho_{AB}) \right] \right)\right\}  =  0  \qquad   \forall x\in  \{1,2,3,4\}\, \forall i\in \{1,
\dots,  N\} \, .
\end{align}
The proof is concluded by observing that  the l.h.s. of this equation is equal to  $\Tr[ P_i  \map T_x  (\rho_i)]$. \qed  

\medskip  

Lemma \ref{lem:telechan} establishes a first connection between bomb-and-prize strategies and teleportation-like schemes were system $Q$ is measured jointly with system $B$, initially correlated with system $A$,  which then undergoes a correction operation.  

In addition, Lemma \ref{lem:telechan}  shows that the states $(  \rho_{i})_{i=1}^N$ can be {\em conclusively excluded} \cite{bandyopadhyay2014conclusive}, that is, that there is a POVM $(Q_i)_{i=1}^N$ such that $\Tr [  Q_i  \rho_i] =  0 \, \forall i$  (indeed, the POVM $(Q_i)_{i=1}^N$ is implicitly defined in Lemma \ref{lem:telechan} and can be made explicit via the relation $\Tr  [  Q_i\, \rho]   =   \Tr [  P_i  \, \map T_x (\rho)]$, $\forall \rho \in  L(\spc H_Q)$.)

We now restrict our attention to the qubit case. In this case, a theorem by Ruskai \cite{Ruskai2003}  guarantees that every entanglement-breaking channel can be represented as a random mixture of classical-to-quantum (CQ) channels,   that is, channels of the form 
\begin{eqnarray}
  \map N_{\rm cq}(\rho) = \sigma_0 \, \langle \psi_0|\rho|\psi_0\rangle + \sigma_1\,  \langle \psi_1|\rho|\psi_1\rangle \, ,
  \label{cqq}
\end{eqnarray}
where $\{|\psi_0\> ,  |\psi_1\>\}$ is an orthonormal basis  and  $\{\sigma_0, \sigma_1\}$ are density matrices.    
Hence, every entanglement-breaking qubit channel $\map N_{\rm eb}$  can be written as
\begin{eqnarray}
\map N_{\rm eb} (\rho)= \sum_{j=1}^{K}   \,q_j  \,   \map N_{\rm cq}^{(j)}(\rho) \, ,
\label{ebcqexpansion}
\end{eqnarray} 
where $K$ is a positive integer,  $(q_j)_{j=1}^K$ is a probability distribution satisfying $q_j>0 \, ,\forall j  \in  \{1,\dots, K\}$, and  
\begin{eqnarray}
  \map N_{\rm cq}^{(j)}(\rho) = \sigma^{(j)}_0 \, \langle \psi^{(j)}_0|\rho|\psi_0^{(j)}\rangle + \sigma^{(j)}_1\,  \langle \psi_1^{(j)}|\rho|\psi_1^{(j)}\rangle \, ,
  \label{cqq}
\end{eqnarray}  
is a CQ channel.  

Combining the convex decomposition (\ref{ebcqexpansion}) with Lemma \ref{lem:telechan}, we obtain the following lemma:  \begin{lemma}\label{lem:telecq}
  If an entanglement-breaking qubit channel $\map N_{\rm eb}$ satisfies  the bomb-avoiding condition (\ref{quantumbombavoid}), then it can be decomposed as 
  \begin{align}\label{quasi}
\map N_{\rm eb}      =    \,  \sum_{j=1}^K q_j  \,     \map N_{\rm cq*}^{(j)}  \, \qquad   {\rm with} \qquad \map N_{\rm cq*}^{(j)}  (\rho)   =  \sum_{m  \in \{0,1\}}   \,   |\phi_{m}^{(j)}\>  \<\phi_{m}^{(j)}| ~~ \<  \psi_{m}^{(j)}| \,\rho\,|\psi_{m}^{(j)}  \> \, ,
\end{align}
where $K$ is a positive integer, $(q_j)_{j=1}^K$ is a probability distribution,  $\{|\psi_{m}^{(j)}\> \}_{m\in  \{0,1\}}$ and  $\{|\phi_{m}^{(j)}\> \}_{m\in  \{0,1\}}$ are orthonormal bases for every $j$, and  
\begin{align}\label{telecq}
\map T_x   (  |\phi_m^{(j)}\>\<\phi_m^{(j)}| )  =  |\psi_{m\oplus 1}^{(j)}  \>\< \psi_{m\oplus 1}^{(j)}|    \, ,\qquad \forall    m\in  \{0,1\}, \, \forall j\in  \{  1,\dots,  K\}, \, \forall x \in  \{1,2,3,4\} \, ,
\end{align}
$\oplus$ denoting the modulo-$2$ addition.   Two CQ channels  $\map N_{\rm cq*}^{(j_1)}$ and $\map N_{\rm cq*}^{(j_2)}$ in Eq. (\ref{quasi}) are distinct if and only if 
\begin{align}
\left\{|\phi_{0}^{(j_1)}\>\<\psi_0^{(j_1)}| \, ,|\phi_{1}^{(j_1)}\>\<\psi_1^{(j_1)}| \right\} \not = \left\{|\phi_{0}^{(j_2)}\>\<\psi_0^{(j_2)}| \, ,|\phi_{1}^{(j_2)}\> \<\psi_1^{(j_2)}|\right\} \, .
\end{align}

\end{lemma}
\Proof Lemma \ref{lem:telechan} implies the condition  
\begin{align}\label{mnb}
 q_j  \,   \<\psi_m^{(j)} |  \,  \map T_x   (    \sigma_m^{(j)}  ) \, |\psi_m^{(j)}\>   =  0  \qquad\forall j\in\{1,
 \dots,  N\},  \,  m\in \{0,1\} \, , \forall x\in  \{1,2,3,4\}   \, ,
\end{align}
as one can see by setting $i=  (j,m)$, $  P_i  =  \,  q_j\,    |\psi_m^{(j)}\>\<\psi_m^{(j)}|$, and $\rho_i  =  \sigma_m^{(j)}$. 

Since $q_j \not  =  0$ for every $j$, Eq.  (\ref{mnb}) implies  that the states $\sigma_0^{(j)}$ and $\sigma_1^{(j)}$ are perfectly distinguishable for every $j$: indeed,  a protocol for distinguishing them is to apply the channel $\map T_x$ and then measure on the basis $\{|\psi_0^{(j)}\>,  |\psi_1^{(j)}\>\}$.    

Hence, $\sigma_0^{(j)}$ and $\sigma_1^{(j)}$ are pure orthogonal states, of the form  $\sigma_0^{(j)}  =  |\phi_0^{(j)}\>\<\phi_0^{(j)}|$ and $\sigma_1^{(j)}  =  |\phi_1^{(j)}\>\<\phi_1^{(j)}|$ for some orthonormal basis $\{|  \phi_0^{(j)}\>,  |\phi_1^{(j)}\>\}$.

  Eq.  (\ref{mnb}) then becomes 
  \begin{align}
  \<\psi_m^{(j)} |  \,  \map T_x   (   |\phi_m^{(j)}\>\<\phi_m^{(j)}|   ) \, |\psi_m^{(j)}\>   =  0  \qquad\forall j\in\{1,
 \dots,  K\},  \,  m\in \{0,1\} \, , \forall x\in  \{1,2,3,4\}   \, ,
  \end{align}
  which is equivalent to Eq. (\ref{telecq}). 

  Finally, if two channels $\map N^{(j_1)}_{\rm cq*}  \not  =  \map N^{(j_2)}_{\rm cq*}$ are distinct, they must have 
  \begin{align}
  \left\{|\phi_{0}^{(j_1)}\>\<\psi_0^{(j_1)}| \, ,|\phi_{1}^{(j_1)}\>\<\psi_1^{(j_1)}| \right\} \not = \left\{|\phi_{0}^{(j_2)}\>\<\psi_0^{(j_2)}| \, ,|\phi_{1}^{(j_2)}\> \<\psi_1^{(j_2)}|\right\} \, .
  \end{align}  
  Indeed, if the  two sets of projectors  were the same, we would have either $|\phi_m^{(j_2)}\>  \propto  |\phi_m^{(j_1)}\> \, ,\forall m\in  \{0,1\}$, or $|\phi_m^{(j_2)}\>  \propto   |\phi_{m\oplus 1}^{(j_1)}\> \, ,\forall m\in  \{0,1\}$.        In both cases, the equality of the two bases  would imply $\map N^{(j_1)}_{\rm cq*}=  \map N^{(j_2)}_{\rm cq*}$, as one can see by using the relation
    \begin{align}\map N^{(j)}_{\rm cq}  (\rho)   =   \sum_m \,    |\phi_{m}^{(j)}\>\<\phi^{(j)}_m|   \,   \Tr \left[ \rho  \,  \map T_x  \left(  |\phi_{m\oplus 1 }^{(j)}\>\<\phi^{(j)}_{m\oplus 1}| \right) \right] \,, \qquad   \forall  m,  \forall j  
    \end{align}
    implied by Eq. (\ref{telecq}). 
    
    Conversely, it is immediate to see that any two CQ channels that prepare states in two distinct bases are necessarily distinct, because one channel outputs states that are diagonal in one basis, and the other output channels that are diagonal in the other basis.   
  \qed

\medskip



We are now ready to establish a correspondence between quantum bomb-and-prize strategies and teleportation schemes: when the entanglement-breaking channel $\map N_{\rm eb}$ is a mixture of two (or more) distinct CQ channels,  every quantum bomb-and-prize strategy using channel $\map N_{\rm eb}$ and satisfying the bomb-avoiding condition is unitarily equivalent to a quantum teleportation scheme. Specifically, the bipartite state $\rho_{AB}$, the measurement $(M_y)_y$ and the encoding operations $(\map E_{b,x})_b$ define a quantum teleportation protocol for every $x$.

\begin{lemma}\label{lem:txunitary}
Let $\map N_{\rm eb}$ be an entanglement-breaking channel  satisfying the bomb-avoiding condition.  If the decomposition (\ref{quasi}) contains  (at least) two distinct CQ channels,   then there exists a unitary channel $\map U:   L  (\spc H_Q) \to L(\spc H_Q)$ such that 
\begin{align}
\map T_x    =  \map U  \, \qquad \forall x\in  \{1,2,3,4\} \, .
\end{align}
\end{lemma}
  \Proof  If  channel $\map N_{\rm eb}$ has a decomposition (\ref{quasi}) containing at least two distinct CQ channels, these two channels must prepare states in two different bases, say 
  \begin{align}
  \left\{|\phi^{(j_1)}_m\>  \< \phi^{(j_1)}_m|  \right\}_{m\in  \{0,1\}}\not =  \left\{|\phi^{(j_2)}_m\> \< \phi^{(j_2)}_m|  \right\}_{m\in  \{0,1\} } 
  \end{align}
  for some $j_1, j_2$  (cf. Lemma \ref{lem:telecq}).      
  
  Now, let $V_x: \spc H_Q  \to \spc H_Q \otimes \spc H_{\rm aux}$   be a Stinespring isometry for channel $\map  T_x$, meaning that  $\map   T_x  (\rho) = \Tr_{\rm aux}  [   V_x \rho  V_x^\dag]$,  where $\rm aux$ is an auxiliary system.    Condition (\ref{telecq}) implies the conditions 
  \begin{align}
      \Tr_{\rm aux}  [V_x   |\phi^{(j)}_m\>\<\phi^{(j)}_m|   V_x^\dag]     =  |\psi^{(j)}_{m\oplus 1}\>\<\psi^{(j)}_{m\oplus 1}|  \, , \qquad \forall j, \forall  m  \, .     
  \end{align}
  In turn, these conditions imply that there exist states $|\xi^{(j,x)}_m\>$  (not necessarily orthogonal), such that 
  \begin{align}\label{stine}
  V_x   \,|\phi^{(j)}_m\>  =  |\psi^{(j)}_{m\oplus 1} \>\otimes |\xi_m^{(j,x)}\>     \qquad \forall j,  \, \forall m \, .  
  \end{align}
  Since the projectors corresponding to the bases $\{|\phi^{(j_1)}_m\>\}_{m\in  \{0,1\}}$ and   $\{|\phi^{(j_2)}_m\>\}_{m\in  \{0,1\} }$ are distinct, we can write $|\psi^{(j_2)}_0\>  =  \alpha\, |\phi^{(j_1)}_0\> + \beta \,  |\phi^{(j_1)}_1\>$, with $\alpha \not  =  0$ and $\beta \not  =  0$.  
Hence, linearity of the operator $V_x$ implies 
\begin{align}
\nonumber  \alpha\, |\phi^{(j_1)}_0\>\otimes  |\xi^{(j_2,x)}_0\> + \beta \,  |\phi^{(j_1)}_1\> \otimes  |\xi^{(j_2,x)}_0\>   &=   |\phi^{(j_2)}_0\>  \otimes |\xi^{(j_2,x)}_0\>   \\
\nonumber &  =  V_x   |\phi^{(j_2)}_0\> \\
\nonumber &=  \alpha\,  V_x   |\phi^{(j_1)}_0\>  +  \beta\,  V_x   |\phi^{(j_1)}_1\> \\
&= \alpha\,   |\phi^{(j_1)}_0\>  \otimes |\xi^{(j_1,x)}_0\>    +  \beta\, |\phi^{(j_1)}_1\>  \otimes |\xi^{(j_1,x)}_1\>   \, .     
\end{align}
  Comparing the first and last term of the above equation, we obtain $|\xi_0^{(j_1,x)}\>  =  |\xi_1^{(j_1,x)}\>  = : |\xi^{(x)}\>$. Inserting this condition into Eq. (\ref{stine}), we obtain the relation 
   \begin{align}
  V_x   \,|\phi^{(j_1)}_m\>  =  |\psi^{(j_1)}_{m\oplus 1} \>\otimes |\xi^{(x)}\>     \qquad  \forall m \, ,  
  \end{align}
which in turn implies $V_x  =  U \otimes |\xi^{(x)}\>$, with $U   :  = \sum_m  |\psi^{(j_1)}_{m\oplus 1} \>\< \phi^{(j_1)}_m |$.  Hence,  we obtain the relation 
\begin{align}
\map T_x  (\rho)    =  \Tr_{\rm aux}   [   V_x \rho  V_x^\dag]    =  U  \rho   U^\dag    =:  \map U(\rho)\qquad  \forall \rho  \in L(\spc H_A) \, ,\forall x\, .    \end{align}
Hence, we conclude that the equality $\map T_x  =  \map U$ holds for every $x$. 
\qed 

\medskip 

Lemma \ref{lem:txunitary} shows that, up to a unitary transformation, the bipartite state $\rho_{AB}$, the measurement $(M_y)_y$ and the encoding operations $(\map E_{b,x})_b$ define a quantum teleportation protocol for every $x$.

\medskip  
The results obtained so far lead us to the following conclusion: 
\begin{lemma}\label{lem:almostthere}
If an entanglement-breaking qubit channel $\map N_{\rm eb}$ satisfies the bomb-avoiding condition, then it is unitarily equivalent to  a {\em $\tt NOT$ channel}, that is, a channel $\map N_{\tt NOT}$  of the form  
\begin{align}\label{done!}
\map N_{\tt NOT}     =    \,  \sum_{j=1}^K q_j  \,   \map N_{\rm cq \, , \tt NOT}^{(j)}   \,  ,   \qquad   \map N_{\rm cq, \, \tt NOT }^{(j)}   (\rho) :  = \sum_{m  \in \{0,1\}}   \,   |\psi_{m\oplus 1}^{(j)}\>  \<\psi_{m\oplus 1}^{(j)}| ~~ \<  \psi_{m}^{(j)}| \,\rho\,|\psi_{m}^{(j)}  \>    \, , \qquad   \qquad \forall \rho \in L(\spc H_Q)  \, ,     
\end{align}
where $K  \ge 1$ is a positive integer, $(q_j)_{j=1}^K$ is a probability distribution with $q_j >  0 \, ,\forall j\in  \{1,\dots,  K\}$, and,  for every $j$,  $\{ |\psi_0^{(j)}\>,  |\psi_1^{(j)}\>\}$ is an orthonormal basis.
\end{lemma}      
\Proof   By Lemma \ref{lem:telecq},  channel $\map N_{\rm eb}$ can be decomposed as  in Eq.  (\ref{quasi}).    If the decomposition (\ref{quasi}) contains only one CQ channel $\map N_{\rm cq*}$  (that is, if $\map N_{\rm cq*}^{(j)}=  \map N_{\rm cq*}$ for every $j$),  then    $\map N_{\rm eb}  (\rho)  = \sum_{m  \in \{0,1\}}   \,   U|\psi_{m\oplus 1}  \>  \<\psi_{m\oplus 1}|U^\dag ~~ \<  \psi_{m}| \,\rho\,|\psi_{m}  \>  \, , \forall \rho$, with  $U:  =\sum_{m\in\{0,1\}}   |\phi_m\>\<\psi_{m\oplus 1}|$.  

Now, let us assume that the  decomposition (\ref{quasi})  contains at least two distinct CQ channels.  Then, Lemma \ref{lem:txunitary}  implies that there exists a unitary channel $\map U$ such that  $\map T_x  =  \map U $ for every $x$.   Inserting this expression into Eq.  (\ref{telecq}),   we then obtain  $\map U(  |\phi^{(j)}_m\>\<\phi^{(j)}_m|)  =  |\psi_{m\oplus 1}^{(j)}\>\<\psi_{m\oplus 1}^{(j)}|\, ,\forall m,j$, which is equivalent to $  |\phi^{(j)}_m\>\<\phi^{(j)}_m|)  = U^\dag |\psi_{m\oplus 1}^{(j)}\>\<\psi_{m\oplus 1}^{(j)}|   U, \, \forall m,j$ where $U$ is the unitary operator corresponding to channel $\map U$.  Hence,  we have $
\map N_{\rm eb}  (\rho)   =  \sum_{j=1}^K q_j  \,  \left( \sum_{m  \in \{0,1\}}   \,   U^\dag |\psi_{m\oplus 1}^{(j)}\>  \<\psi_{m\oplus 1}^{(j)}|U ~~ \<  \psi_{m}^{(j)}| \,\rho\,|\psi_{m}^{(j)}  \>\right) \, ,\forall \rho$.  \qed

\medskip

Lemma \ref{lem:almostthere} shows that every entanglement-breaking qubit channel satisfying  the bomb-avoiding condition must be unitarily equivalent to a $\tt NOT$  channel, that is, a channel of the form (\ref{done!}).      It remains to show the converse part:  if a channel is unitarily equivalent to a $\tt NOT$ channel, then it satisfies the bomb-avoiding condition. This proof uses the following lemmas:
\begin{lemma}\label{lem:degradetounot}
  Every $\tt NOT$ channel   $\map N_{\tt NOT}$     can be stochastically degraded to the $\tt UNOT$ channel, that is, there exists  a nonzero probability $\lambda>0$ and a qubit channel $\map C$ such that $  \lambda\, \map N_{\rm eb}   +  (1-\lambda) \,  \map C   =  \tt UNOT$.    
\end{lemma}

\Proof  Suppose that $\map N$   is a $\tt NOT$ channel.    Each CQ channel $N_{\rm cq, \, \tt NOT}^{(j)}$ in Eq. (\ref{done!}) can be equivalently written as $
\map N_{\rm cq, \, \tt NOT}^{(j)}   =  \frac 12  \,  (\map V^{(j)}_+  +  \map V^{(j)}_-)$, where $\map V^{(j)}_{\pm}$  are the unitary channels corresponding to the unitary operators $V^{(j)}_\pm: = |\psi_1^{(j)}\> \<\psi_0^{(j)}|  \pm |\psi_0^{(j)}\> \<\psi_1^{(j)}|$. 
   Since both operators are traceless and orthogonal with respect to the Hilbert-Schmidt product, they are unitarily equivalent to the Pauli matrices $X$ and $Y$, say $ V^{(j)}_+ =   R_j  X   R_j^\dag  $ and  $R_j  Y   R_j^\dag$, for some unitary operator $R_j$.  Defining the unitary operator $W_j: =  R_j  Z R_j^\dag  $ and the corresponding unitary channel $\map W$,  we have 
   \begin{align}
\nonumber      \frac2 3  \,  \map N_{\rm cq, \, \tt NOT}^{(j)}  (\rho)    +  \frac 13  \,   \map W_j   (\rho) &=   \frac{ R_j  X   R_j^\dag  \rho  R_j  X   R_j^\dag   +  R_j  Y   R_j^\dag  \rho  R_j  Y   R_j^\dag  +R_j  Z   R_j^\dag  \rho  R_j  Z   R_j^\dag    }3\\
\nonumber & =R_j     \Big (  {\tt UNOT}    \left(R_j^\dag \rho  R_j \right) \Big)  \,  R_j^\dag  \\
  &  =  {\tt UNOT}  (\rho)   \, \qquad \forall \rho  \in  L (\spc H_Q) \,, 
   \end{align}
where the last equality follows form the fact that the $\tt UNOT$ channel is unitarily covariant  \cite{wernerbuzek}.  Hence, every CQ channel $\map N^{(j)}_{\rm cq,\, \tt NOT}$ in Eq. (\ref{done!}) can be degraded to the $\tt UNOT$ channel.  Since every $\tt NOT$ channel is a  convex combination of CQ channels that can be degraded to the $\tt UNOT$ channel, it can also  be degraded to the $\tt UNOT$ channel.

\qed

\medskip

 \begin{lemma}\label{lem:degradeimpliesavoid}
  If a qubit channel $\map N$ can be stochastically degraded to the $\tt UNOT$ channel, then it  satisfies the bomb-avoiding condition.       
 \end{lemma}
 \Proof   Let us write  ${\tt UNOT}  =  \lambda  \map N  +  (1-\lambda)  \map C$ for some nonzero probability $\lambda>0$ and some qubit channel $\map C$.  Then, for every fixed state $ \rho_{AB}$, every fixed set of encoding operations $( \map E_{b,x} )_{b,x}$, and every fixed measurement $(  M_y)_y$, we can compare the two bomb-and-prize  protocols using channels $\map N$  and $\tt UNOT$, respectively.    The probability that Bob finds outcome $y$ when the bomb is in $b$ and the prize is in $x$, given by Eq. (\ref{pybxquantum})  satisfies  
the inequality  
\begin{align}
\lambda\,  p_{  \map N}  (y| b,x)  \le p_{ \tt UNOT } (y| b,x)  \qquad \forall b\in  \{1,2,3,4\}  \, , \forall x\in  \{1,2,3,4\}\, .
\end{align}  
In particular, let set $y=  b$.  Since the probability $\lambda$ is non-zero,  the condition $p_{  \tt UNOT } (b| b,x)  =  0,  \, \forall b,x$ implies the condition $p_{  \map N_{\tt NOT} } (b| b,x)  =  0,  \,  \forall b,x$.   Since the $\tt UNOT$ channel satisfies the bomb-avoiding condition in the $\tt UNOT$-dense-coding protocol, then also channel $\map N$ satisfies the bomb-avoiding condition.    \qed  

\medskip

\begin{lemma}\label{lem:notimpliesavoid}
  If a qubit channel $\map N_{\rm eb}$ is unitarily equivalent to a $\tt NOT$ channel, then it satisfies the bomb-avoiding condition.       
 \end{lemma}

 \Proof    Let us write $\map N_{\rm eb}  =  \map U  \circ \map N_{\tt NOT}  \circ \map V$ where $\map N_{\tt NOT}$ is a $\tt NOT$ channel, and $\map U$ and $\map V$ are two unitary channels.   Then,   $\map N_{\tt NOT}$ can be stochastically degraded to the $\tt UNOT$ channel (by Lemma \ref{lem:degradetounot}), and satisfies the bomb-avoiding condition (by Lemma \ref{lem:degradeimpliesavoid}).  Since $\map N_{\rm eb}$ is unitarily equivalent to a channel that satisfies the bomb-avoiding condition, it also satisfies the bomb-avoiding condition. \qed

\medskip

Lemmas \ref{lem:almostthere} and \ref{lem:notimpliesavoid}  prove the first sentence in the statement of Theorem 3:  an entanglement-breaking channel satisfies the bomb-avoiding condition if and only if it is unitarily equivalent to a $\tt NOT$ channel.   Before moving to the rest of the Theorem,  we show that the results proven so far give us a complete characterization of the $\tt NOT$ channels:  

\begin{proposition}\label{prop:notchannels}
   A qubit channel  $\map N$  is a $\tt NOT$ channel if and only if it is entanglement-breaking and  can be stochastically degraded to the $\tt UNOT$ channel.  
\end{proposition}
\Proof 
Suppose that $\map N$ is a $\tt NOT$ channel. Then, it is entanglement-breaking by definition.   Moreover, Lemma \ref{lem:degradetounot} implies that $\map N$ can be degraded to the $\tt UNOT$ channel.  

Conversely, suppose that $\map N$ is entanglement-breaking and can be degraded to the $\tt UNOT$ channel.   Then, Lemma  \ref{lem:degradeimpliesavoid} implies that $\map N$ satisfies the bomb-avoiding condition, and Lemma \ref{lem:almostthere} implies that $\map N$ is unitarily equivalent to a $\tt NOT$ channel.   In particular, $\map N$ can be decomposed as $\map N   =  \sum_j \,  q_j  \, \map N_{\rm cq*}^{(j)}$ with $\map N_{\rm cq*}^{(j)}  (\rho)   =  \sum_{m\in  \{0,1\}}   \,  |\phi_m^{(j)}\>\<\phi_m^{(j)}|   \, \<  \psi_m^{(j)}|  \rho  |\psi_m^{j}\> \, , \forall \rho$, where $\{  |\psi_0^{(j)}  \>,\, |\psi_1^{(j)}  \>\}$ and $\{  |\phi_0^{(j)}  \>,\, |\phi_1^{(j)}  \>\}$ are two orthonormal bases. 

Now, the degradation  condition   ${\tt UNOT}  =  \lambda  \map N  +  (1-\lambda)  \map C$, with some nonzero probability $\lambda>0$ and some qubit channel $\map C$,    implies 
\begin{align}
\nonumber 0  &= \< \Phi_+|   ({\tt UNOT} \otimes \map I)  \, (|\Phi^+\>\<\Phi^+|) \, |\Phi^+\> \\
\nonumber &\ge  \lambda\,  \< \Phi_+|   (\map N \otimes \map I)  \, (|\Phi^+\>\<\Phi^+|) \, |\Phi^+\> \\
\nonumber &  = \lambda\,  \sum_j \, q_j \< \Phi_+|   (\map N_{\rm cq *}^{(j)} \otimes \map I)  \, (|\Phi^+\>\<\Phi^+|) \, |\Phi^+\> \\
&  =  \lambda\,  \sum_j\,  q_j  \,   \left(  \sum_{  m\in\{ 0,1 \}}  \,  \frac{|\<  \phi_m^{(j)}|\psi_m^{(j)}\>|^2}2\right)  \,    
\end{align}
Hence, we must have $|\<  \phi_m^{(j)}|\psi_m^{(j)}\>|  =  0$ for every $m$ and for every $j$ such that $q_j\not = 0$.    In other words, every CQ channel $\map N_{\rm cq*}^{(j)}$ appearing with nonzero probability in the decomposition of $\map N$ must be a $\tt NOT$ channel, and therefore also $\map N$ is a $\tt NOT$ channel.  \qed 
\medskip

  The remainder of  Theorem 3 is about {\em quantum {\tt NOT} channels}, that is, channels of the form (\ref{done!}), where the sum contains two or more distinct terms. 
These channels can be characterized as follows: 
\begin{proposition}\label{prop:quantumnotchannels}
For a qubit channel $\map N$, the following are equivalent:  
\begin{enumerate}
\item $\map N$  a quantum $\tt NOT$ channel 
\item $\map N$ is entanglement-breaking and it is equivalent to the $\tt UNOT$ channel under stochastic degradation, that is, if and only if   there exist two nonzero probabilities $\lambda, \mu  >0$, and two qubit channels $\map C,  \map D$ such that ${\tt UNOT}   =  \lambda\,   \map N_{\rm eb}  +  (1-\lambda) \,  \map C $  and $\map N_{\rm eb}   =  \mu\,   {\tt UNOT}  +  (1-\mu) \,  \map D $. 
\end{enumerate}
\end{proposition}

    \Proof 
    $1\Rightarrow 2.$  Suppose that $\map N$ is a quantum $\tt NOT$ channel.  Then, $\map N$ is entanglement-breaking and  Proposition \ref{prop:notchannels} ensures that  it can be stochastically degraded to the $\tt UNOT$ channel.  It remains to show that the $\tt UNOT$ channel can be stochastically degraded to $\map N$.   Since $\map N$ is a quantum $\tt NOT$ channel, it has a decomposition   (\ref{done!}) containing at least two distinct CQ channels $\map N_{\rm cq*}^{(j_1)}$ and $\map N_{\rm cq*}^{(j_2)}$,  corresponding to two different bases,  $\{ |\psi_0^{(j_1)}\>,  |\psi_1^{(j_1)}\>\}$ and  $\{ |\psi_0^{(j_2)}\>,  |\psi_1^{(j_2)}\>\}$, respectively.     Since these two bases correspond to distinct observables, there must exist two nonzero complex numbers $\alpha\not  = 0$ and $\beta  \not = 0$ such that   
\begin{align}
\nonumber |\psi_0^{(j_2)}  \>     & =   \alpha  \,   |\psi_0^{(j_1)}  \>     +\beta  \,  |\psi_1^{(j_1)}  \>  \\
|\psi_1^{(j_2)}  \>     & \propto    \overline \beta  \,   |\psi_0^{(j_1)}  \>    - \overline \alpha  \,  |\psi_1^{(j_1)}  \> \, .
\end{align}
Defining $V^{(j)}_\pm: = |\psi_1^{(j)}\> \<\psi_0^{(j)}|  \pm |\psi_0^{(j)}\> \<\psi_1^{(j)}|$ for $j\in  \{j_1,j_2\}$, we then have 
\begin{align}
\nonumber V_+^{(j_2)} &\propto    \left[   -\alpha^2   \,|\psi_0^{(j_1)}\> \<\psi_1^{(j_1)}|   + \beta^2 \, |\psi_1^{(j_1)}\> \<\psi_0^{(j_1)}|   + \alpha \beta \, \left( |\psi_0^{(j_1)}\> \<\psi_0^{(j_1)}|   -   |\psi_1^{(j_1)}\> \<\psi_1^{(j_1)}| \right) \right]   +   {\rm Hermitian~conjugate}\\    
\nonumber &=   {\sf Re}  [-\alpha^2 + \beta^2]   \, \left( |\psi_0^{(j_1)}\> \<\psi_1^{(j_1)}|     +   |\psi_1^{(j_1)}\> \<\psi_0^{(j_1)}| \right)    \\ 
\nonumber & \qquad  +   i  {\sf Im}  [-\alpha^2 + \beta^2]      \, \left( |\psi_0^{(j_1)}\> \<\psi_1^{(j_1)}|   -   |\psi_1^{(j_1)}\> \<\psi_0^{(j_1)}| \right) \\     
\nonumber &    \qquad +  2 {\sf Re}  [\alpha \beta] \, \left( |\psi_0^{(j_1)}\> \<\psi_0^{(j_1)}|   -   |\psi_1^{(j_1)}\> \<\psi_1^{(j_1)}| \right)\\
  &  = {\sf Re}  [-\alpha^2 + \beta^2]  \,  X^{(j_1)}   -     {\sf Im}  [-\alpha^2 + \beta^2]   \, Y^{(j_1)}  + 2 {\sf Re}  [\alpha \beta] \, Z^{(j_1)}  \,,
\end{align}
having defined 
\begin{align}
\nonumber    X^{(j_1)}  &: =  |\psi_0^{(j_1)}\> \<\psi_1^{(j_1)}|     +   |\psi_1^{(j_1)}\> \<\psi_0^{(j_1)}|  \equiv  V^{(j_1)}_+\\
\nonumber Y^{(j_1)}   &: =    -i |\psi_0^{(j_1)}\> \<\psi_1^{(j_1)}|     +i   |\psi_1^{(j_1)}\> \<\psi_0^{(j_1)}|       \equiv  i\,  V^{(j_1)}_-    \\
 Z^{(j_1)}  &: =  |\psi_0^{(j_1)}\> \<\psi_0^{(j_1)}|     -   |\psi_1^{(j_1)}\> \<\psi_1^{(j_1)}|  \,.    
\end{align}
Similarly, we obtain  
\begin{align}
\nonumber V_-^{(j_2)} &\propto    \left[   -\alpha^2   \,|\psi_0^{(j_1)}\> \<\psi_1^{(j_1)}|   + \beta^2 \, |\psi_1^{(j_1)}\> \<\psi_0^{(j_1)}|   + \alpha \beta \, \left( |\psi_0^{(j_1)}\> \<\psi_0^{(j_1)}|   -   |\psi_1^{(j_1)}\> \<\psi_1^{(j_1)}| \right) \right]   -   {\rm Hermitian~conjugate}\\    
\nonumber &=   {\sf Re}  [-\alpha^2 + \beta^2]   \, \left( |\psi_0^{(j_1)}\> \<\psi_1^{(j_1)}|     -   |\psi_1^{(j_1)}\> \<\psi_0^{(j_1)}| \right)    \\ 
\nonumber & \qquad  +   i  {\sf Im}  [-\alpha^2 + \beta^2]      \, \left( |\psi_0^{(j_1)}\> \<\psi_1^{(j_1)}|   +   |\psi_1^{(j_1)}\> \<\psi_0^{(j_1)}| \right) \\     
\nonumber &    \qquad +  2i {\sf Im}  [\alpha \beta] \, \left( |\psi_0^{(j_1)}\> \<\psi_0^{(j_1)}|   -   |\psi_1^{(j_1)}\> \<\psi_1^{(j_1)}| \right)\\
& =  i {\sf Re}  [-\alpha^2 + \beta^2]  \,  Y^{(j_1)}   +    i   {\sf Im}  [-\alpha^2 + \beta^2]   \, X^{(j_1)}  + 2i {\sf Im}  [\alpha \beta] \, Z^{(j_1)}  \,,   \,.
\end{align}
The above equations imply the relation 
\begin{align}\label{linspan}
\Span  \{     V^{(j_1)}_+  , \,     V^{(j_1)}_-, \,   V^{(j_2)}_+  , \,   V^{(j_2)}_-\}= \Span   \{    X^{(j_1)}  \, , Y^{(j_1)}   \, , Z^{(j_1)}  \} \, .
\end{align}

To conclude, we  use the Choi representation.    For a linear map $\map M:    L(\spc H_Q) \to L(\spc H_Q)$, let  $\Choi   (\map M) :  =  ( \map M \otimes \map I_Q)  ( \Phi^+\>\<\Phi^+|)$ be the corresponding Choi operator \cite{Choi1975}.   In the Choi representation, Eq. (\ref{linspan})  implies that the support of the Choi operator $  \Choi  ( q_{j_1} \,  \map N_{\rm cq}^{(j_1)}   +  q_{j_2} \,  \map N_{\rm cq}^{(j_2)} )$  contains the support of the operator  
\begin{align}
\nonumber &\frac{   (  X^{(j_1)} \otimes I_Q)\,   |\Phi^+\>\<\Phi^+|   \,   (  X^{(j_1)} \otimes I_Q)^\dag  +  (  Y^{(j_1)} \otimes I_Q)\,   |\Phi^+\>\<\Phi^+|   \,   (  Y^{(j_1)} \otimes I_Q)^\dag  +  (  Z^{(j_1)} \otimes I_Q)\,   |\Phi^+\>\<\Phi^+|   \,   (  Z^{(j_1)} \otimes I_Q)^\dag }3  \\
& \qquad \equiv  \Choi  ({\tt UNOT}) \, .
\end{align}
Hence, we have the inclusion $\Supp   \left(  \Choi  (\map N) \right) \supseteq  \Supp \left(   \Choi( q_{j_1} \,  \map N_{\rm cq}^{(j_1)}   +  q_{j_2} \,  \map N_{\rm cq}^{(j_2)} ) \right) \supset \Supp (  \Choi  ({  \tt UNOT}))$.  Hence, there exists a strictly positive number $\mu>0$ such that 
\begin{align}
\mu \,  \Choi  ({\tt UNOT})  \le    \Choi  (\map N_{\rm eb})  \,,    
\end{align}
or equivalently,  
\begin{align}
    \Choi  (\map N) =   \mu \,  \Choi  ({\tt UNOT})  +  Q    \,,    
\end{align}
where $Q$ is some positive operator.  
Taking the partial trace of the above equation on the first Hilbert space, we then obtain $I/2   =  \mu  \, (I/2)  +  \Tr_1  [ D ]$, which implies $\Tr_1  [D]   =   (1-\mu) \,  I/2$.  Hence, $D$ is proportional to the Choi operator of a  quantum channel, with proportionality constant $(1-\mu)$.  Writing $D  =  (1-\mu)\, \Choi  (\map D)$, we then obtain   
\begin{align}
    \Choi  (\map N) =   \mu \,  \Choi  ({\tt UNOT})  + (1-\mu)\,  \Choi  (\map D)    \,,    
\end{align}
or equivalently, 
\begin{align}
\map N   =  \mu  \,  {\tt UNOT}   +  (1-\mu )\, \map D \, .
\end{align}

$2\Rightarrow 1.$ Suppose that $\map N$ is entanglement-breaking and equivalent to the $\tt UNOT$ channel under stochastic degradation. Then, Proposition \ref{prop:notchannels} implies that $\map N$ is a $\tt NOT$ channel.  It remains to show that $\map N$ is a quantum $\tt NOT$ channel, namely that is decomposition (\ref{done!}) contains at least two distinct CQ channels.  This fact can be easily shown by contrapositive. Suppose that the decomposition of $\map N$ contains a single CQ channel, say $\map N(\rho)   = \sum_{m\in\{0,1\}}   |\psi_{m\oplus 1}\>\<\psi_{m\oplus 1}|  \,  \<\psi_m|  \rho |\psi_m\>  \, , \forall \rho$, for some basis $\{|\psi_0\>, |\psi_1\>\}$, and suppose that  $\map N  =  \mu \,  {\tt UNOT}  +  (1-\mu)\,  \map D$ for some probability $\mu  \ge 0$ and some qubit channel $\map D$.  Then, we have 
\begin{align}
\nonumber 0&=  \<\psi_0|  \map N (|\psi_0\>\<\psi_0|) |\psi_0\> \\
\nonumber &\ge \mu\,  \<  \psi_0  |  \,  {\tt UNOT} (|\psi_0\>\<\psi_0|) \, |\psi_0\>    \\
\nonumber  
&  =    \mu\,  \<  \psi_0  |  \,   \left(  \frac 23  I  - \frac 13 |\psi_0\>\<\psi_0|  \right) \, |\psi_0\> \\
&  =  \frac \mu 3 \,    
\end{align}
which implies $\mu  =  0$.    Hence, if the $\tt UNOT$ channel can be stochastically degraded to a $\tt NOT$ channel $\map N$, then  $\map N$ must be a quantum $\tt NOT$ channel.   \qed

\medskip  

\begin{lemma}\label{lem:c<1}
For a qubit channel $\map N$, the following are equivalent:  
\begin{enumerate}
\item $\map N$ is an entanglement-breaking channel with non-unit entanglement-assisted capacity $C_{\rm E}  (\map N)  < 1$ and satisfies the bomb-avoiding condition
\item $\map N$ is unitarily equivalent to a quantum $\tt NOT$ channel.
\end{enumerate}
\end{lemma}

\Proof  $1\Rightarrow 2$.   Since $\map N$ is entanglement-breaking and satisfies the bomb-avoiding condition, Lemma  \ref{lem:almostthere} ensures  that  $\map N$  is unitarily equivalent to a $\tt NOT$ channel, say $\map N  =  \map U \circ \map N_{\tt NOT}  \circ \map V$, where where $\map U$ and $\map V$ are unitary channels, and $\map N_{\tt NOT}$ is a $\tt NOT$ channel.
 Now, consider  the decomposition (\ref{done!}) of channel  $\map N_{\tt NOT}$. If this decomposition  contains only one CQ channel,  then the classical capacity  (and therefore the entanglement-assisted capacity) of channel $\map N$ is $1$.   Hence, for the capacity to be less than $1$, the decomposition (\ref{done!}) must contain at least two distinct CQ channels, meaning that $ \map N_{\tt NOT}$  is a quantum $\tt NOT$ channel.

$2\Rightarrow 1$.  Since $\map N$ is unitarily equivalent a  $\tt NOT$ channel, it is  entanglement-breaking. Moreover,  Lemma  \ref{lem:notimpliesavoid} guarantees that $\map N$ satisfies the bomb-avoiding condition.    Now, let us write $\map N$ as $\map N  =  \map U\circ \map N_{\tt qNOT}  \circ \map V$ where $\map U$ and $\map V$ are unitary channels and $\map N_{\tt qNOT}$ is a quantum $\tt NOT$ channel.  Then, we have 
\begin{align}
 C_{\rm E}  (\map N)  =  C_{\rm E}  (\map N_{\tt qNOT}) \, .   
\end{align}  
Now, Proposition \ref{prop:quantumnotchannels}  implies that $\map N_{\tt qNOT}$ can be decomposed as   $\map N_{\tt qNOT}  =  \mu\,  {\tt UNOT} +  (1-\mu)\,  \map D$, for some non-zero probability $\mu>0$ and some qubit channel $\map D$. Then,   the convexity of the entanglement-assisted capacity implies the bound 
\begin{align}
\nonumber C_{\rm E}  (\map N_{\tt qNOT})  &\le  \mu\,    C_{\rm E}  ({\tt UNOT})+ (1-\mu)\,   C_{\rm E}   (\map D)  \\
 &  \le   \mu \,  (  2-  \log 3)   +  (1-\mu)\,   C_{\rm E}   (\map D) \, .
\label{c<1}
\end{align}
Since $\map D$ can stochastically degraded to $\map N_{\tt qNOT}$ and $\map N_{\tt qNOT}$ can be stochastically degraded to $\tt UNOT$  (by Proposition \ref{prop:quantumnotchannels}), we have that $\map D$ can be stochastically degraded to $\tt UNOT$.  Hence, Proposition \ref{prop:notchannels} implies that $\map D$ is a $\tt NOT$ channel, and, in particular, is entanglement-breaking. Then, $C_{\rm E}  (\map D)  \le 1$, and Eq. (\ref{c<1}) becomes $C_{\rm E}  (\map N_{\rm eb})   \le  \mu  (2 -\log 3)  +  (1-\mu)   <  1$. 

  \qed

The above lemma proves the second sentence of Theorem 3: an entanglement-breakining channel with nonunit entanglement-assisted capacity satisfies the bomb-avoiding condition if and only if it is a quantum $\tt NOT$ channel.

\section{Proof of Theorem 4}

We start by showing that, for every protocol using a  quantum $\tt NOT$ channel,  the minimum amount of entanglement is one ebit, corresponding to two maximally entangled qubits: 

\begin{lemma}\label{lem:atleast1ebit}
   Every  bomb-avoiding protocol using a  quantum $\tt NOT$ channel  requires a correlated state $\rho_{AB}$ that is at least as entangled as a two-qubit maximally entangled state ({\em i.e.}    $\rho_{AB}$ can be converted into a maximally entangled two-qubit state using local operations and classical communication.)
\end{lemma}
\Proof    Lemma \ref{lem:telecq} implies that every bomb-avoiding protocol using a quantum $\tt NOT$ channel  is associated to a quantum teleportation protocol that uses the state $\rho_{AB}$ as the initial resource shared by Alice and Bob.     It is well known \cite{bennett1993teleporting,werner2001all} that every teleportation protocol for a qubit requires an initial state $\rho_{AB}$ that is as entangled as the two-qubit maximally entangled state $|\Phi^+\>$.  \qed 

\medskip  

The above lemma shows   the dense-coding protocols introduced in  the main text require the minimum amount of entanglement.   We now analyze their performance in terms of the worst-case probability of finding the prize.

\begin{lemma}\label{lem:prize}
Every bomb-avoiding protocol using a quantum $\tt NOT$ channel   and a two-qubit correlated state satisfies the condition $p_{\rm worst}^{\rm prize}  \le 1/3$.   
\end{lemma}

\Proof For a protocol using two-qubit entangled states and a quantum $\tt NOT$ channel,   the encoding operations $\map E_{b,x}$ must be independent of $x$. Indeed, Lemma \ref{lem:txunitary} implies that the channels 
\begin{align}
\map T_x  (\rho)   :=  \sum_b  \,    \map E_{b,x} \left(  \Tr_{QB}  \Big[(  M_b\otimes I_A)   \, ( \rho \otimes \rho_{AB} )  \Big] \,   \right)  
\end{align}
are all equal to the same unitary channel.  Since systems $Q, A,$ and $B$ are all qubits, this condition implies that the quantum operation $\map Q_b$ defined by
\begin{align}
\map Q_b  (\rho) :  = \Tr_{QB}  \Big[(  M_b\otimes I_A)   \, ( \rho \otimes \rho_{AB} )  \Big] , \,  \qquad \forall \rho  \in  L(\spc H_Q) 
\end{align}
is proportional to a unitary channel $\map U_b$ and the encoding channel $\map E_{b,x}$ is equal to its inverse $\map U_b^{-1}$.  Hence, the encoding $\map E_{b,x}$ is independent of $x$, the position of the prize.  This fact implies that the probability $p_{\map N}(  y|  b,x)$ in Eq. (\ref{pybxquantum}) is independent of $x$.  In particular, there exists a probability distribution $q  (x|b)$ such that  
\begin{align}
p_{\map N}(  x|  b,x)  =   q (x|b)  \qquad \forall b,x \, .     
\end{align}
Hence, $p_{\rm worst}^{\rm prize}  =  \min_{b,x}  q  (x|b)  \le 1/3$, the inequality following from the relation $q(b|b)  =  p_{\map N}  (b|b,x)  =  0$, implied by the bomb-avoiding condition. 

 \qed

\medskip

\section{Proof of Theorem 5}  
\label{app:th3}

We now show that simulating the 4-dimensional {\tt NOT} channel through the communication of one bit requires Alice and Bob to share correlated classical systems of dimension $d\ge 3$.  Moreover, we show that every simulation using systems of dimension $d=3$ requires the maximum amount of shared randomness, equal to $\log 3$.  In other words,  this simulation requires Alice and Bob  to share a pair of $3$-dimensional systems that are perfectly correlated and have uniformly random marginals.  

 The proof  is divided into 5 parts: 
\begin{enumerate}
\item In the first part, we formalize the problem of simulating  the $4$-dimensional {\tt NOT} channel with minimum amounts of shared randomness.  Mathematically, the problem amounts to finding an optimal convex decomposition of the  $4$-dimensional {\tt NOT} channel into channels  that can be implemented without shared randomness. 
\item  In the second part, we show that, without loss of generality, the minimization of the shared randomness can be restricted to strategies where Alice and Bob use perfectly correlated systems.  
\item In the third part,  we show that the search for optimal strategies can be restricted to a set of strategies that we call off-diagonal. 
\item In the fourth part, we classify the off-diagonal strategies achievable by Alice and Bob without shared randomness.
\item In the  fifth part,  we show that simulating the $4$-dimensional {\tt NOT} channel requires a mixture of at least three strategies, and that the probabilities in the mixture must be uniform, corresponding $\log 3$ bits of shared randomness.  
\end{enumerate}

\subsection{Problem formulation} 


Here we formulate the problem of minimizing the amount of correlations needed to simulate the $4$-dimensional {\tt NOT} channel, setting the notation used in the rest of the section.    

We  start by formalizing the possible strategies available to Alice and Bob.  
Without shared randomness, the most general strategy consists of an encoding channel, described by the probability  $p_{\rm enc}    (m|  b)$ that Alice sends the message $m\in  \{0,1\}$ when the input variable is $b\in  \{1,2,3,4\}$, and a decoding channel, described by the probability  $p_{\rm dec}  (y|m)$ that Bob's output variable is $y \in  \{ 1,2,3,4\}$ upon receiving the message $m$.

In the following, strategies   with no shared randomness will be represented by encoding-decoding pairs, of the form  $\map S=     (p_{\rm enc}, p_{\rm dec})$.    For a strategy   without shared randomness, the distribution of the input and output variables is 
\begin{align}
p_{\map S} :  =     \sum_{m\in  \{0,1\}}   \,    p_{\rm dec}  (y|m) \,  p_{\rm enc}  (m|b) \, .
\end{align}

Now, suppose that Alice and Bob share two correlated classical systems in the states $i$ and $j$, respectively.  In this case, the encoding channel can depend on $i$, and the corresponding probability distribution will be denoted by  $p_{\rm enc}^{(i)}  (m|  b)$.  Similarly, the decoding channel can depend on $j$, and the corresponding probability distribution will be denoted by  $p_{\rm dec}^{(j)}  (y|  m)$.   The joint probability of $i$ and $j$ will be denoted by  $p_{i j}$.  
Without loss of generality, we will assume that Alice's and Bob's systems have the same dimension, hereafter denoted by $d$.

In the following, strategies with shared randomness will be denoted by tuples of the form  $  \map R    =  \left(p_{ij} \,  ,  p_{\rm enc}^{(i)},  p_{\rm dec}^{(j)}   \right)_{i,  j  \in  \{1,\dots, d\} } $.      For the strategy  $\map R$, the input-output probability distribution is  
\begin{align}
p_{\map R}  (y|b)   :=  \sum_{i,j=1}^d \,       p_{ij}   \, 
\sum_{m  \in  \{0,1\}}  \,     p_{\rm dec}^{(j)}  (y|m)  \,  p_{\rm enc}^{(i)}   (m|b)    \, .
\end{align}

\begin{definition}
We say that a strategy   $  \map R $  {\em simulates the $4$-dimensional {\tt NOT} channel} if 
\begin{align}\label{simulatenot}
p_{\map R}  (y|b)     =  p_{\tt NOT}   (y|b)  \equiv     \frac 13\,     (1-  \delta_{y,b}) \qquad \forall y,b  \in  \{1,2,3,4\} \, .
\end{align}  
\end{definition}

Among the strategies that simulate the $4$-dimensional {\tt NOT} channel, our goal is to find a strategy that minimizes the amount of randomness shared between Alice and Bob, which can be quantified by the mutual information  of the random variables $i$ and $j$. Explicitly, the mutual information is 
\begin{align}
\nonumber H(A:B)    &  =  H(  A)   +  H  (B)  -  H(A,B) \\
&  =  -  \sum_{i=1}^d    q_i  \log  q_i     -      \sum_{j=1}^d    r_j  \log  r_j +     \sum_{i=1}^d  \sum_{j =1}^d    p_{ij}  \log  p_{ij}       \, ,  \label{mutualinfo}
\end{align}  
with  $q_i : =  \sum_j   p_{ij}$ and $r_j: =  \sum_i   p_{ij}$.   In the following we will address the minimization of  the mutual information (\ref{mutualinfo}) over all strategies $\map R$  satisfying  the simulation constraint (\ref{simulatenot}).  

\subsection{Reduction to  perfectly correlated strategies}

The search for the optimal simulation of the $4$-dimensional {\tt NOT} channel is simplified by the following lemma: 
\begin{lemma}\label{lem:perfectcorrelation}
For every fixed dimension of Alice's and Bob's input systems, the search for a strategy that  simulates the $4$-dimensional {\tt NOT} channel and  minimizes the mutual information can be restricted without loss of generality to strategies  $  \map R    =   \Big(p_{ij},     p^{(i)}_{\rm enc} ,  p^{(j)}_{\rm dec}  \Big)_{i,j\in \{1,\dots, d\}}$   where  Alice's and Bob's system are perfectly correlated, namely $p_{ij}  =  \delta_{i j}   q_i$, $\forall i \, ,\forall j$.
\end{lemma}
 
 \Proof   The idea is that every lack of correlation between Alice's and Bob's system can be reabsorbed in the definition of the encoding and decoding operations.

 Suppose that  there exists one value $j_0$ and two values $i_1$ and $ i_2$ such that $p_{i_1  j_0}  >  0$ and  $p_{i_2   j_0}  >  0$.  In this case, one can replace the probability distribution $ (p_{ij})_{i,j} $ with the new probability distribution  $ (\widetilde p_{ij})_{i,j} $ defined as
 \begin{align} 
 \widetilde p_{ij}    := 
  \begin{cases}
p_{i_1 j_0}   +  p_{ i_2   j_0}    &   {\rm for}~(i,j)    =  (i_1,  j_0) \\
0   &   {\rm for}~(i,j)    =  ( i_2,  j_0) \\
p_{ij}   & {\rm otherwise}
 \end{cases} 
 \end{align}
 and one can replace Alice's encoding  $p_{\rm enc}^{(i)}  (m|b)$  with the new encoding  
\begin{align}
\widetilde p_{\rm enc}^{(i)}  (m| b):  = 
\begin{cases}
\frac{p_{i_1 j_0} }{p_{i_2  j_0}   +  p_{i_2   j_0}  }   \,   p_{\rm enc}^{(i_1)}  (m|  b )  +   \frac{p_{i_2  j_0 } }{p_{i_1  j_0}   +  p_{i_2  j_0}  }   \,   p_{\rm enc}^{(i_2)}  (m|  b  )  \qquad  &\qquad {\rm for}~  i  = i_1 \\
p_{\rm enc}^{(i)}  (m|b)    &\qquad {\rm otherwise}\,.
\end{cases}
\end{align}
These replacements do not alter the input-output distribution $p(y|b)$, and therefore produce a new simulation of the $4$-dimensional {\tt NOT} channel.  On the other hand, since the the probability distribution $  (\widetilde p_{ij})_{i,j} $ is obtained from   $ (p_{ij})_{i,j}  $  by post-processing  Alice's variable, its mutual information cannot be larger than the mutual information of $(p_{ij})_{i,j}$ due to the data processing inequality.  

By iterating the above procedure, we then obtain a new simulation strategy satisfying the condition that for every $j$ there exists at most  one value of $i$ such that $p_{  i,j} >  0$.  Now, suppose that, for this new simulation strategy,   there exists one value $i_0$ and two values $j_1$ and $j_2$ such that $p_{i_0  j_1}  >  0$ and  $p_{i_0  j_2}  >  0$.  In this case, one can apply the same replacement approach to Bob's decoding:  one can replace the probability distribution $p_{ij}$ with the new probability distribution 
 \begin{align} 
 \widetilde p_{ij}  := 
  \begin{cases}
p_{i_0 j_1}   +  p_{ i_0 j_2}   &   {\rm for}~(i,j)    =  (i_0,j_1) \\
0   &   {\rm for}~(i,j)    =  (i_0, j_2) \\
p_{ij}   & {\rm otherwise}
 \end{cases} 
 \end{align}
and  Bob's decoding  $p_{\rm dec}^{(j)}  (y|m)$  with the new decoding  
\begin{align}
\widetilde p_{\rm dec}^{(j)} (y|m):  = 
\begin{cases}
\frac{p_{i_0 j_1} }{p_{ i_0j_1}   +  p_{i_0  j_2}  }   \,   p_{\rm dec}^{(j_1)}  (y|  m  )  +   \frac{p_{i_0  j_1} }{p_{i_0  j_1}   +  p_{i_0 j_2}  }   \,   p_{\rm dec}^{(j_1)}  (y|  m  )  \qquad &\qquad {\rm for}~   j =  j_1 \\
 \\
p_{\rm dec}^{(j)}  (y|m)   \qquad &\qquad {\rm otherwise}\,.
\end{cases}
\end{align}
Again,  these replacements do not alter the input-output distribution $p(y|b)$, and therefore produce a new simulation of the $4$-dimensional {\tt NOT} channel.  On the other hand, since the the probability distribution $ (\widetilde p_{ij})_{ij}$ is obtained from   $ p_{ij}  $  by post-processing  Bob's variable, its mutual information cannot be larger than the mutual information of $p_{ij}$.  

By iterating the above procedure, we then obtain a new simulation strategy satisfying the condition that for every value of $i$ with $q_i  >  0$ there exists one and only one value of $j$ such that $p_{ij}  >  0$.  In other words,  there exists a permutation $\pi$ such that $p_{ij}  =  q_i   \,  \delta_{j,  \pi  (i)}$.  At this point,  the states of Bob's system can be relabelled in such a way that they are perfectly correlated with the states of Alice's system.  \qed

\medskip

Thanks to  Lemma \ref{lem:perfectcorrelation}, we can restrict our attention to strategies with  perfectly correlated probability distributions  $p_{ij}  =  q_i  \,  \delta_{ij}$.   Without loss of generality, we will assume that 
\begin{align}\label{ps>0}
q_i  >  0 \, \qquad \forall i\in  \{1,\dots, d\}  
\end{align}  
(if this condition is not satisfied, we can always reduce the value of $d$ until the condition is satisfied.)

 In the following,  a strategy with perfect correlations will be described by a  tuple  $\map R=  \Big(q_i,  p_{\rm enc}^{(i)},  p^{(i)}_{\rm dec} \Big)_{i=1,\dots,  d}$.  For strategy $\map R$,    the input-output distribution is
 \begin{align}\label{pybdecomp}
 p_{\map R}(y|b)    =    \sum_i   \,  q_i   \,  p_{\map S_i}   (y|b) \, ,    
 \end{align}   
 where 
 \begin{align}\label{pSi} p_{\map S_i}     (y|b) :  =\sum_{m\in  \{0,1\}}   \,  p_{\rm dec}^{(i)}  (y   |    m)\,  p^{(i)}_{\rm enc}    (m|   b )
 \end{align} 
 is the input-output distribution of strategy $\map S_i   =  \left (p_{\rm enc}^{(i)},   p_{\rm dec}^{(i)} \right)$. 
    
The simulation constraint (\ref{simulatenot}) amounts to the condition that the probability distributions   $ p_{\map S_i}  (y|b)$ in Eq. (\ref{pSi}) provide a convex decomposition of the $4$-dimensional {\tt NOT} channel into strategies without shared randomness,  namely  
\begin{align}\label{simulatenot1}
\sum_i   \,  q_i  \,     p_{ \map S_i}   (  y|b)  =  \frac 13 \,  (1-\delta_{y,b})  \qquad \forall y,b \in  \{1,2,3,4\} \, .
\end{align}

For strategies with perfect correlations, the mutual information $H(A:B)$ in Eq. (\ref{mutualinfo}) reduces to the Shannon entropy of the probability distribution $  {\bf q}  : =  (q_i)_{i\in \{1,\dots, d\}}$, namely
 \begin{align}\label{mutualinfo1}
 H  (A:B)   =   - \sum_{i=1}^d  \,  q_i  \log   q_i  =:  H(   {\bf q}  )\, .
 \end{align} 
 Our goal will be to minimize  the Shannon entropy  $H({\bf q})$  over all  convex combinations $  \sum_i   \,  q_i   \,     p_{ \map S_i}   (  y|b)$ that satisfy  the simulation constraint (\ref{simulatenot1}).


\subsection{Reduction to off-diagonal strategies}

For the conclusion of our proof of necessity,  It is convenient to use the following matrix representation: 
\begin{definition}\label{def}
 The {\em strategy matrix} of  a strategy with input-output distribution $p(y|b)$ is the matrix $M$  with entries  $M_{y,b} := p(y|b)$.
 \end{definition}

Note that the simulation constraint  (\ref{simulatenot}) implies the condition 
\begin{align}\label{bombavoid}
M_{b,b}   =0  \qquad\forall b\in  \{1,2,3,4\}  \, .
\end{align}
 
 \begin{definition}
 We say that a strategy is {\em off-diagonal} if the corresponding  strategy matrix $M$   satisfies Eq. (\ref{bombavoid}).  
 \end{definition}
  Off-diagonal strategies form a convex set,   which contains in particular  all the simulations of the $4$-dimensional {\tt NOT} channel.   

\begin{lemma}
Let  $\map R=  \Big(q_i,  p_{\rm enc}^{(i)},  p^{(i)}_{\rm dec} \Big)_{i=1,\dots,  d}$ be a perfectly correlated strategy with shared randomness, and, for every $i \in  \{1,\dots, d\}$, let    $\map S_i   :=  \Big  (p_{\rm enc}^{(i)},  p^{(i)}_{\rm dec} \Big)$. If $\map R$ is off-diagonal, then $\map S_i$ is off-diagonal  for every $i$.  
\end{lemma}  
 \Proof Immediate from the relation $0   =  p_{\map R}(  b|b)  = \sum_i\,  q_i  \,  p_{\map S_i}  (b|b)\, , \forall b$ and from the fact that $q_i$ is strictly positive for every $i$.   \qed

\subsection{Classification of the strategy matrices of off-diagonal strategies without shared randomness}
               
We now show that  off-diagonal strategies  without shared randomness  can be divided into four types.  To this purpose, we first show that every off-diagonal strategy without shared randomness is associated to a  pair of disjoint, nonempty subsets.  
\begin{lemma}\label{lem:disjoint}
Let  $ \map S=   \left(p_{\rm enc} ,   p_{\rm dec}   \right) $ be a strategy without shared randomness, and   let 
 $B_0$ and $B_1$ be the sets defined by 
\begin{align}\label{Bm}
B_m  :  =      \,\{   y  \,  |     p_{\rm dec}    (y|  m)   > 0   \}  \qquad \forall  m \in  \{0,1\}\, .
\end{align} 
If $\map S$ is off-diagonal, then  $B_0  \cap B_1 = \emptyset$.    

\end{lemma}
 
 \Proof  The proof is by contrapositive. Suppose that  the intersection  $B_0  \cap  B_1$  is non-empty,   and let $b_*$ be an element of $ B_0  \cap  B_1$.    Then, 
 \begin{align}\label{insomma}
  p_{\rm dec}     (b_*|  m)   > 0 \, ,  \qquad \forall  m\in \{0,1\} \,.
  \end{align}  Hence, one has
 \begin{align}
\nonumber  M_{b_*,  b_*}  &  \equiv p(b_*|b_*)    \\
\nonumber     &  =    \sum_m  \,  \,  
 p_{\rm dec}  (b_*|  m)   \,  p_{\rm enc}   (m|  b)   \\
   \nonumber   &  \ge      p_{\rm dec}    (b_*|  0)  \,  p_{\rm enc}   (0| b_* )     +    p_{\rm dec}   (b_*|  1) \,  p_{\rm enc}  (1| b_* )      \Big]\\
   \nonumber &  \ge      \max\Big\{  p_{\rm dec}  (b_*|  0)  \,  p_{\rm enc} (0| b_* )      \,  , \,      p_{\rm dec}   (b_*|  1) \,  p_{\rm enc} (1| b_* )   \Big\}  \\ 
      \nonumber &  \ge       \min  \Big\{  p_{\rm dec}  (b_*|  0)        ,     p_{\rm dec}  (b_*|  1)     \Big\}  ~  \max \Big\{   p_{\rm enc} (0| b_* )   , \,    p_{\rm enc}   (1| b_* )  \Big\}  \\ 
     &  >0\, ,  \label{uffa}
 \end{align} 
 where the last inequality follows from the bounds  $\max\Big\{   p_{\rm enc}   (0| b_* )   , \,    p_{\rm enc}   (1| b_* )  \Big\}  >  0$ (implied by the normalization condition  $p_{\rm enc} (0| b_* )   +   p_{\rm enc}  (1| b_*) =1$),   and    $\min  \Big\{  p_{\rm dec}   (b_*|  0)        ,     p_{\rm dec}  (b_*|  1)     \Big\}  >0$ (implied by Eq.  (\ref{insomma}) .)    
Eq. (\ref{uffa}) shows that condition (\ref{bombavoid}) is not satisfied.  By contrapositive, if condition (\ref{bombavoid})  is satisfied, then  the intersection $ B_0 \cap  B_1$  must be empty.    \qed  

\medskip

Thanks to Lemma \ref{lem:disjoint}, off-diagonal strategies without share randomness  are associated to pairs of disjoint nonempty subsets $(B_0, B_1)$.   In fact, the order of the subsets is irrelevant, because the labels $0$ and $1$ can be exchanged without affecting the overall input-output distribution:  
\begin{lemma}\label{lem:flip}
Let  $ \map S=   \left(p_{\rm enc} ,   p_{\rm dec}   \right) $ be a strategy without shared randomness, and   let 
 $B_0$ and $B_1$ be the sets defined in Eq. (\ref{Bm}).   Then, there exists another off-diagonal strategy without shared randomness, denoted by $\map S'$, such that $p_{\map S'}   (y|b)   =  p_{\map S}  (y|b)  \, , \forall y,  b$, and $B_0'  =  B_1$ and $B_1'  =  B_0$, where $B_0'$ and $B_1'$ are the sets associated to $\map S'$ as in Eq. (\ref{Bm}). 
 \end{lemma}

\Proof  Define $p_{\rm enc} ' (m|b)   =  p_{\rm enc}   (m\oplus 1|  b) $ and  $p_{\rm dec} ' (y|m)   =  p_{\rm dec}   (y|  m\oplus 1)$.   Then, the strategy $\map S'  :  =  \left(p_{\rm enc}',  p_{\rm dec}'  \right)$ has the desired properties. \qed  

\medskip

Using this fact, every off-diagonal  strategy without shared randomness can be classified by an unordered pair  $\{ B_0,  B_1  \}$, with $B_m  \subset  \{1,2,3,4\} \, \forall m\in \{0,1\}$,  $B_0  \not  =  \emptyset,  \,  B_1 \not =  \emptyset,  B_0  \cap  B_1  =  \emptyset$.  
As a convenient  bookkeeping tool, we introduce the following definition:  
\begin{definition}
Let $\map S  = \left( p_{\rm enc},  p_{\rm dec} \right)$  be an off-diagonal strategy without shared randomness, and let $B_0$ and $B_1$ be the associated sets defined in Eq. (\ref{Bm}).  Then,  
  we say that 
  \begin{itemize}
  \item $\map S$ is {\em of type $(m,n)$}, with $m:=  \min    \{|B_0|,  |B_1|  \}$  and $n:=  \max \{|B_0|,   |B_1|\}$, and 
  \item    $\map S$ is {\em of subtype} $\{B_0,  B_1\}$.     
   \end{itemize}  
  \end{definition}

Since  $B_0$ and $B_1$ are nonempty disjoint subsets of $\{1,2,3,4\}$,  there are four possible types:  $  (1,1)$,   $(1,2)$,  $(1,3)$, and $(2,2)$.       
We now analyze the form of  the  matrices associated to strategies of these four possible types.

\medskip  

{\em   Type $(1,1).$}   In this case, there are $6$ subtypes:  indeed, the unordered pair $\{B_0, B_1\}$ can be  $\big\{\{1\}   ,  \{2\}\big\}$,    $\big\{\{1\}   ,  \{3\}\big\}$,    $\big\{\{1\}   ,  \{4\}\big\}$,   $\big\{\{2\}   ,  \{3\}\big\}$,   $\big\{\{2\}   ,  \{4\}\big\}$,  or $\big\{\{3\}   ,  \{4\}\big\}$.

\begin{proposition}\label{prop:1}   
Let $\map S  =  (   p_{\rm enc},  p_{\rm dec})$  be a strategy of type $(1,1)$.   If $\map S$ is of subtype    $\big\{\{1\}   ,  \{2\}\big\}$,    $\big\{\{3\}   ,  \{4\}\big\}$,    $\big\{\{1\}   ,  \{3\}\big\}$,    $\big\{\{2\}   ,  \{4\}\big\}$,   $\big\{\{1\}   ,  \{4\}\big\}$,   or $\big\{\{2\}   ,  \{3\}\big\}$,      then its strategy matrix is of the form
\begin{align}
\nonumber   &M_{1|2}    =  
\begin{pmatrix}
0         &  1   &  p   &   r \\
1         &  0   &  1-p   &  1-r \\
0         & 0   &  0   & 0 \\
0           & 0   &  0   & 0 
\end{pmatrix}   \, ,  \qquad 
M_{3|4}    =  
\begin{pmatrix}
0         &  0   &  0   &   0 \\
0         &  0   &  0   &    0 \\
p         &   r &  0   & 1 \\
1-p           & 1-r   &  1   & 0 
\end{pmatrix} \\
\nonumber &M_{1|3}    =  
\begin{pmatrix}
0         &  p   &  1   &   r \\
0         &  0   &  0   &    0 \\
1         &  1-p  &  0   & 1-r \\
0           & 0   &  0   & 0 
\end{pmatrix}   \, ,  \qquad 
M_{2|4}    =  
\begin{pmatrix}
0         &    0   &  0   &   0 \\
p         &  0   &   r   &    1 \\
0         &   0 &    0  &   0 \\
1-p       & 1   &  1-r   & 0 
\end{pmatrix}  \\
&M_{1|4}    =  
\begin{pmatrix}
0         &  p   &  r  &   1 \\
0         &  0   &  0   &    0 \\
0         &   0  &    0  &   0 \\
1           & 1-p   &  1-r   & 0 
\end{pmatrix} \, ,  
\qquad
M_{2|3}    =  
\begin{pmatrix}
0         &  0   &  0   &   0 \\
p         &  0   &  1   &    r\\
1-p         & 1   &  0  &   1-r \\
0          &   0   &  0   & 0 
\end{pmatrix}  
\end{align}   
respectively, where $p$ and $r$ are arbitrary probabilities.    Note that all  matrices of type $(1,1)$ have two rows of zeros.  
 \end{proposition}

\Proof  We provide the proof in the case $B_0   =   \{1\}$ and $B_1 =  \{  2\}$, as the argument is the same in all cases.  
    Since $B_0$ contains 1,  the input value $b=1$ cannot encoded into the message $m=0$, for otherwise the off-diagonal condition $p(1|1)= 0$ would be violated. Hence, $b=1$ is encoded into message $m=1$, and, since $B_1  =  \{1\}$ the output value can only be $y=2$.  In short, we have proven that $  p(2|1)  = 1$.  Since the columns of the strategy matrix contains probabilities normalized to one, we conclude that the first column of matrix $M_{1|2}$ has a $1$ in position $(2,1)$ and zeros everywhere else.    In the same way, we can show that the second column has a $1$ in position $(1,2)$ and zeros everywhere else.     Now, consider the input  value $b=3$.  Since this value is neither contained in $B_0$ nor in $B_1$, it can be encoded into both messages $m=0$ and $m=1$, with probabilities $ p_{\rm enc}   (0|3) =:  p$ and      $ p_{\rm enc}   (1|3) =:1-  p$, respectively. In these two cases, the output value will be either $y=1$ or $y=2$, respectively.   Hence, the third column of matrix $M_{1|2}$ has $p$ in position $(1,3)$,  $1-p$ in position $(2,3)$, and zeros everywhere else. The same argument shows that the fourth column has a probability $r$ in position $(1,4)$, $1-r$ in position $(2,4)$, and zeros everywhere else.  \qed 
    
 \medskip

 {\em   Type $(1,2)$.}   In this case, there are $12$ subtypes: indeed, the unordered pair $\{B_0, B_1\}$ can be  $\big\{\{1\}   ,  \{2,3\}\big\}$,    $\big\{\{1\}   ,  \{2,4\}\big\}$,    $\big\{\{1\} ,  \{3,4\}\big\}$,  $\big\{\{2\} ,  \{1,3\}\big\}$,  $\big\{\{2\} ,  \{1,4\}\big\}$,   $\big\{\{2\} ,  \{3,4\}\big\}$,   $\big\{\{3\} ,  \{1,2\}\big\}$,  $\big\{\{3\} ,  \{1,4\}\big\}$,   $\big\{\{3\} ,  \{2,4\}\big\}$,    $\big\{\{4\} ,  \{1,2\}\big\}$,    $\big\{\{4\} ,  \{1,3\}\big\}$,   $\big\{\{4\} ,  \{2,3\}\big\}$. 

 \begin{proposition}\label{prop:3}    
Let $\map S  =  (   p_{\rm enc},  p_{\rm dec})$  be an off-diagonal strategy without shared randomness.  If the corresponding pair of subsets is $\{B_0, B_1\}$ can be  $\big\{\{1\}   ,  \{2,3\}\big\}$,    $\big\{\{1\}   ,  \{2,4\}\big\}$,    $\big\{\{1\} ,  \{3,4\}\big\}$,  $\big\{\{2\} ,  \{1,3\}\big\}$,  $\big\{\{2\} ,  \{1,4\}\big\}$,   $\big\{\{2\} ,  \{3,4\}\big\}$,   $\big\{\{3\} ,  \{1,2\}\big\}$,  $\big\{\{3\} ,  \{1,4\}\big\}$,   $\big\{\{3\} ,  \{2,4\}\big\}$,    $\big\{\{4\} ,  \{1,2\}\big\}$,    $\big\{\{4\} ,  \{1,3\}\big\}$,   $\big\{\{4\} ,  \{2,3\}\big\}$,    then the strategy matrix is of the form
\begin{align}
\nonumber   &M_{1|23}    =  
\begin{pmatrix}
0         &  1   &  1   &   1-r \\
p         &  0   &  0   &    rp\\
1-p         &   0   &  0   &  r(1-p) \\
0           & 0   &  0   & 0 
\end{pmatrix}   ,  \, 
M_{1|24}    =  
\begin{pmatrix}
0         &   1  &  1-r   &   1 \\
p         &  0   &  rp  &    0 \\
0         &   0 &  0   &      0 \\
1-p           & 0   &  r(1-p)   & 0 
\end{pmatrix} , \,  
M_{1|34}    =  
\begin{pmatrix}
0         &  1-r   &   1   &   1 \\
0         &  0   &  0   &    0 \\
p         &   rp  &  0   & 0 \\
1-p           & r(1-p)   &  0   &  0   
\end{pmatrix}  , \\
\nonumber   &M_{2|13}    =  
\begin{pmatrix}
0         &  p   &  0   &   rp \\
1         &  0   &  1   &    1-r\\
0       &  1-p   &  0   &  r(1-p) \\
0           & 0   &  0   & 0 
\end{pmatrix}   ,  \, 
M_{2|14}    =  
\begin{pmatrix}
0         &  p   &  rp   &   0 \\
1         &  0   &  1-r   &    1\\
0       &  0  &  0  &  0 \\
0           & 1-p   &  r(1-p)   & 0 
\end{pmatrix}   ,  \, 
M_{2|34}    =  
\begin{pmatrix}
0         &     0  &   0   &   0 \\
1-r         &  0   &  1   &    1 \\
rp         &   p  &  0   & 0 \\
r(1-p)           & 1-p   &  0   &  0   
\end{pmatrix} , \\
\nonumber   &M_{3|12}    =  
\begin{pmatrix}
0         &  0   &  p   &   rp \\
0         &  0   &  1-p   &    r(1-p)\\
1          &  1   &  0   &    1-r \\
0          & 0   &  0   & 0 
\end{pmatrix}   ,  \, 
M_{3|14}    =  
\begin{pmatrix}
0         &  rp   &  p   &   0 \\
0         &  0   &   0   &    0\\
1       &  1-r   &   0   &  1 \\
0           & r(1-p)   &  1-p   & 0 
\end{pmatrix}   ,  \, 
M_{3|24}    =  
\begin{pmatrix}
0         &     0  &   0   &   0 \\
rp         &  0   &  p   &    0 \\
1-r         &   1  &  0   &    1 \\
r(1-p)      & 0   &  1-p   &  0   
\end{pmatrix},  \\
&M_{4|12}    =  
\begin{pmatrix}
0         &  0   &  rp   &   p \\
0         &  0   &  r(1-p)   &   1-p\\
0          &  0  &  0   &    0 \\
1          & 1   &  1-r   & 0 
\end{pmatrix}   ,  \, 
M_{4|13}    =  
\begin{pmatrix}
0         &  rp   &  0   &   p \\
0         &  0   &   0   &    0\\
0       &  r(1-p)   &   0 &  1-p \\
1           & 1-r   &  1   & 0 
\end{pmatrix}   ,  \, 
M_{4|23}    =  
\begin{pmatrix}
0         &     0  &   0   &    0 \\
rp         &  0   &  0   &    p \\
r(1-p)         &   0  &  0   &    1-p \\
1-r      & 1   &  1   &  0   
\end{pmatrix}   ,
\end{align}   
where $p$ and $r$ are arbitrary probabilities.    Note that all matrices of type $(1,2)$ have at least one row of zeros.  
\end{proposition}
Here and in the following we omit the proof, which follows the same arguments of the proof of Proposition \ref{prop:1}.

\medskip 

{\em   Type $(1,3)$. }   In this case, there are $4$ subtypes: indeed, the unordered pair $\{B_0, B_1\}$ can be  $\big\{\{1\}   ,  \{2,3,4\}\big\}$,    $\big\{\{2\}   ,  \{1,3,4\}\big\}$,    $\big\{\{3\} ,  \{1,2,4\}\big\}$,  $\big\{\{4\} ,  \{1,2,3\}\big\}$.
 
\begin{proposition}\label{prop:4}    
Let $\map S  =  (   p_{\rm enc},  p_{\rm dec})$  be an off-diagonal strategy without shared randomness.  If the corresponding pair is   $\big\{\{1\}   ,  \{2,3,4\}\big\}$,    $\big\{\{2\}   ,  \{1,3,4\}\big\}$,    $\big\{\{3\} ,  \{1,2,4\}\big\}$,  $\big\{\{4\} ,  \{1,2,3\}\big\}$, then $M$ is of the form
\begin{align}
\nonumber M_{1|234}    &=  
\begin{pmatrix}
0              &  1   &  1   & 1 \\
p            &  0   &  0   & 0 \\
r            & 0   &  0   & 0 \\
1-r-p       & 0   &  0   & 0 
\end{pmatrix}   \, ,  \qquad   \phantom{\rm or} \qquad
M_{2|134}    =  
\begin{pmatrix}
  0  &  p                     &  0   & 0 \\
  1  &         0      &  1   & 1    \\
 0   &  r            & 0   &  0    \\
  0&   1-p-r       & 0   &  0  
\end{pmatrix}    \, ,\\
\nonumber &  \\
M_{3|124}    & =  
\begin{pmatrix}
  0  &   0  &    p                     &  0   \\
  0  &   0&       r    &  0    \\
  1  &   1  &       0           & 1     \\
  0&     0   & 1-p-r       & 0      
\end{pmatrix} \, ,    \qquad  {\rm or} \qquad
M_{4|123}    =  
\begin{pmatrix}
  0  &   0  &    0&   p         \\
  0  &   0&      0& r         \\
  0  &   0  &     0&   1-p-r       \\
  1&     1   &  1  &    0           
\end{pmatrix} \, ,  \label{matrices}
\end{align}
where $p\ge0,  r\ge 0$, and $p+r\le1$. 
\end{proposition}

  {\em   Type $(2,2)$.}   In this case, there are $3$ possible subtypes: indeed, the unordered pair $\{B_0, B_1\}$ can be  $\big\{\{1,2\}   ,  \{3,4\}\big\}$,    $\big\{\{1,3\}   ,  \{2,4\}\big\}$,   or   $\big\{\{1,4\}  , \{2,3\}\big\}$. 

 \begin{proposition}\label{prop:2}    
Let $\map S  =  (   p_{\rm enc},  p_{\rm dec})$  be a strategy of type $(2,2)$.    If $\map S$ is of subtype   $\big\{\{1,2\}   ,  \{3,4\}\big\}$,    $\big\{\{1,3\}   ,  \{2,4\}\big\}$,   or  $\big\{\{1,4\} , \{2,3\}\big\}$,      then its matrix is of the form
\begin{align}
\nonumber   &M_{12|34}    =  
\begin{pmatrix}
0         &  0   &  r   &   r \\
0         &  0   &  1-r   &  1-r \\
p         &   p   &  0   & 0 \\
1-p           & 1-p   &  0   & 0 
\end{pmatrix}  , \, 
M_{13|24}    =  
\begin{pmatrix}
0         &  r   &  0   &   r \\
p         &  0   &  p   &    0 \\
0         &   1-r  &  0   & 1-r \\
1-p           & 0   &  1-p   & 0  
\end{pmatrix} , \, 
M_{14|23}    =  
\begin{pmatrix}
0         &   r  &  r   &   0 \\
p         &  0   &  0   &    p \\
1-p         &  0  &  0   & 1-p \\
0           & 1-r   &  1-r   & 0 
\end{pmatrix}  ,
\end{align}   
respectively, where $p$ and $r$ are arbitrary probabilities.  
 \end{proposition}

\medskip  

Before concluding the section, we use the above characterization to prove a few useful properties.   

\begin{corollary}\label{lem:offdiag1}
If a strategy is of type $(1,1)$,  $(1,2)$, or $(1,3)$, then its strategy matrix contains at least a $1$ in some off-diagonal position. 
\end{corollary}
 
 \Proof Immediate from Propositions \ref{prop:1}, \ref{prop:3} and \ref{prop:4}. \qed  
 
 \medskip  
 \begin{corollary}\label{lem:offdiag0}
Let $\map S$ and $\map S'$ be two off-diagonal strategies without shared randomness. If $\map S$ is of type $(2,2)$ and $\map S'$ is of a type other than $(2,2)$, then the corresponding strategy matrices $M$ and $M'$ have at least one $0$ in the same off-diagonal position.   
\end{corollary}

\Proof The proof follows by comparison  of the matrices in  Proposition  \ref{prop:2}  with the matrices in Propositions \ref{prop:1}, \ref{prop:3},  and  \ref{prop:4}.  
  \qed 

\medskip

 \begin{corollary}\label{cor:0000}
Let $M$ be a strategy matrix of type $(2,2)$.  If a row of $M$ contains two zeros  in two distinct off-diagonal places, then $M$ contains a $1$ in some off-diagonal position.   
\end{corollary}

\Proof Proposition \ref{prop:2} shows that, for every strategy matrix of type $(2,2)$, the $k$-th row   contains a zero in position $kk$, another zero in an off-diagonal position, and two other off-diagonal entries, equal to some probability $\alpha \in  [0,1]$. If two off-diagonal entries are equal to zero, then necessarily $\alpha=  0$.  But Proposition \ref{prop:2} shows that there is a row containing the entry $1-\alpha$. Hence,  the matrix $M$ contains an entry equal to $1$. \qed

\subsection{Necessity of perfectly correlated, uniformly random trits} 

Here we prove that simulating the $4$-dimensional  {\tt NOT} channel with one bit of classical communication  requires correlated classical  systems of dimension  at least $3$,  and and that every simulation using $3$-dimensional systems requires maximal correlations, corresponding to a mutual information of $\log 3$.  

Mathematically, a  strategy using correlated trits  is described by a convex combination of three off-diagonal strategies $\map S_0$, $\map S_1$, and $\map S_2$, mixed with probabilities $q_0$, $q_1$, and $q_2$,  respectively.     In terms of the corresponding strategy matrices,  the simulation condition (\ref{simulatenot1}) reads
\begin{align}\label{simulatenot2}
\sum_{i\in  \{0,1,2\}}\,  q_i  \,   M_i     =  M_{\tt NOT} \qquad {\rm with} \qquad 
M_{\tt NOT}  : = \begin{pmatrix}
0   &  \frac 13  &  \frac 13   &\frac 13  \\
\frac 13   & 0  & \frac 13  & \frac 13 \\
\frac 13  &  \frac 13  &  0  &  \frac 13 \\
\frac 13  &  \frac 13  &  \frac 13 &  0  
\end{pmatrix} \, .
\end{align}    
In the following, we  will show that condition  (\ref{simulatenot2}) implies that the  probability distribution in the convex combination must be uniform, that is, $q_0=q_1=q_2$.  In particular, this result implies that simulating the $4$-dimensional {\tt NOT} channel requires correlated systems of dimensions at least $3$.  

  For the proof, we will use the following observations: 
 
\medskip  

\begin{lemma}\label{cor:lessthan13}
Let $a,b,$ and $c$ be three indices such that $\{a,b,c\}  =  \{0,1,2\}$, and let $M_a,  M_b,  M_c$ be three strategy matrices such that the simulation condition (\ref{simulatenot2}) is satisfied for some probabilities $q_a,q_b,$ and $q_c$, respectively.     If $M_{a}$ is of type $(1,1)$, $(1,2)$, or $(1,3)$, then  $q_{a}\le 1/3$.  
\end{lemma} 

\Proof By Corollary \ref{lem:offdiag1} there exists an off-diagonal position $kl$  ($k\not =  l$)     such that $ [M_a]_{kl}    = 1$.   Then, the simulation condition  (\ref{simulatenot2}) implies   
    \begin{align} 
   \frac 13   &=  [M_{\tt NOT}]_{kl}   =    \sum_{i\in  \{0,1,2\}}   q_i \,  [M_i]_{kl}   \ge  q_{a}  \, ,
    \end{align}
    the inequality following from the fact that all probabilities (and therefore all matrix elements of strategy matrices) are nonnegative numbers. 
\qed

\medskip  

\begin{lemma}\label{cor:morethan13}
Let $a,b,$ and $c$ be three indices such that $\{a,b,c\}  =  \{0,1,2\}$, and let $M_a,  M_b,  M_c$ be three strategy matrices such that the simulation condition (\ref{simulatenot2}) is satisfied for some probabilities $q_a,q_b,$ and $q_c$, respectively. If matrix $M_a$ is of type $(2,2)$ and matrix $M_b$ is not of type $(2,2)$, then  there exists an off-diagonal position $kl$ such that  $q_c  \,  [M_c]_{kl}  =  1/3$, and therefore $q_c  \ge 1/3$.  
\end{lemma}

\Proof  Corollary \ref{lem:offdiag0}   guarantees that $M_a$ and $M_b$   have at least one $0$ in the same off-diagonal position.    Let $kl$ be an off-diagonal position such that  $[M_a]_{kl}  =  [M_b]_{kl}  = 0$. Hence, the simulation condition (\ref{simulatenot2}) implies  
\begin{align}\label{aaa}
 \frac 1 3  = [M_{\tt NOT}]_{kl}   =  q_c\,  [  M_c]_{kl} \, ,
\end{align}  
which in turn implies $q_c  \ge 1/3$.  \qed

\medskip

We are now ready to prove the main result of this subsection. Specifically, we will prove that {\em (i)}   it is impossible to simulate the $4$-dimensional {\tt NOT} channel with a bit and two correlated bits, and {\em {(ii)}} simulating the $4$-dimensional {\tt NOT} channel with a bit of classical communication and two correlated trits requires the two trits to have uniformly random marginals.  In other words, the mutual information between the two trits must have the maximal value $\log 3$. 


\begin{proposition}
Let  $\map S_0$,  $\map S_1,$ and $\map S_2$  be three off-diagonal strategies without shared randomness, and let $M_0$,  $M_1$, and $M_2$ be the corresponding strategy matrices.  If the simulation condition (\ref{simulatenot2}) is  satisfied with probabilities $q_0, q_1, $ and $q_2$, then $q_0=q_1=q_2=  1/3$.  
\end{proposition}

\Proof   There are four possible cases, depending on the number of strategies of type $(2,2)$.  

\medskip  

{\em Case 1:  none of the strategies is of type $(2,2)$.}  In this case, Corollary \ref{cor:lessthan13} implies  $q_i \le 1/3$, $\forall i\in  \{0,1,2\}$.    Hence,  the normalization condition $\sum_i q_i=1$ implies $q_0=q_1=q_2=\frac 13$.

\medskip  

{\em Case $2$: exactly  one   strategy is of type $(2,2)$.}  Without loss of generality, let us assume that strategy $\map S_0$ is of type $(2,2)$, while strategies $\map S_1$ and $\map S_2$  are of other types.   Then, Corollary \ref{cor:lessthan13} implies  $q_1\le 1/3$ and $q_2\le 1/3$.  Moreover, applying Corollary \ref{cor:morethan13} with $a=0$ and $b=1$ yields $q_2  \ge 1/3$, and  applying Corollary \ref{cor:morethan13} with $a=0$ and $b=2$ yields $q_1  \ge 1/3$   Hence,  we have  $q_1=  q_2  = 1/3$. Normalization of the probability distribution then yields  $q_0  = 1/3$.

\medskip 

{\em Case 3:  exactly two   strategies are of type $(2,2)$.}   Without loss of generality, we assume that strategies $\map S_0$ and $\map S_1$ are of type $(2,2)$, while strategy  $\map S_2$ is of another  type.  

Lemma \ref{cor:morethan13} implies $q_0  \ge 1/3$  (by setting $a=1$, $b=2$, $c=0$) and $q_1  \ge 1/3$  (by setting $a=1$, $b=2$,  $c=0$).  

Moreover,   Lemma \ref{cor:morethan13} with $a=1$, $b=2$, and $c=0$,  implies that  there exists an off-diagonal position $kl$ such that $q_0  \,  [M_0]_{kl}  =  1/3$.    Since  $M_0$ is a matrix of type $(2,2)$, and every row of a matrix of type $(2,2)$ has at least two  off-diagonal entries with the same value,   there exists an index $l'  \not = l$ such that $[M_0]_{kl'}  =  [M_0]_{kl}$.  But then $q_0  \,  [M_0]_{kl'}=  q_0  \,  [M_0]_{kl}  =  1/3$.       
Then, the simulation condition (\ref{simulatenot2})  implies   $q_1 \,  [M_1]_{kl} =  q_1\,  [M_1]_{kl'}  =  0$, and, since $q_1\ge 1/3$, $  [M_1]_{kl} =   [M_1]_{kl'}  =  0  $.    

Since $M_1$ has two zeros in off-diagonal positions,  Corollary \ref{cor:0000}  implies that $M_1$ has at least one off-diagonal entry equal to $1$.      Then,  the simulation condition (\ref{simulatenot2}) implies  $q_1  \le 1/3$.  
Since we already proved the inequality $q_1  \ge 1/3$, we conclude that $q_1  =1/3$.

Iterating the above argument  with  $a=0$, $b=2$, and $c=1$, we  obtain $q_0 = 1/3$.  Finally,  the normalization of the probability distribution  implies $q_2  =  1/3$.  

\medskip  
 
{\em  Case 4: all three strategies are of type $(2,2)$. } Without loss of generality, we assume that the  three strategies are of different subtypes (if two strategies were of the same subtype, then they could be merged into a single strategy of that subtype), and we take $M_1$ to of subtype $\big\{\{1,2\},  \{3,4\}\big\}$,  $M_2$ of subtype $\big\{\{1,3\},  \{2,4\}\big\}$, and   $M_3$ of subtype $\big\{\{1,4\},  \{2,3\}\big\}$.  Using Proposition \ref{prop:2}, we can write these matrices as  
\begin{align}   &M_0    =  
\begin{pmatrix}
0         &  0   &  r_0   &   r_0 \\
0         &  0   &  1-r_0   &  1-r_0 \\
p_0         &   p_0   &  0   & 0 \\
1-p_0           & 1-p_0   &  0   & 0 
\end{pmatrix}  , \, 
M_1    =  
\begin{pmatrix}
0         &  r_1   &  0   &   r_1 \\
p_1         &  0   &  p_1   &    0 \\
0         &   1-r_1  &  0   & 1-r_1 \\
1-p_1           & 0   &  1-p_1   & 0  
\end{pmatrix} , \, 
M_2   =  
\begin{pmatrix}
0         &   r_2  &  r_2   &   0 \\
p_2         &  0   &  0   &    p_2 \\
1-p_2         &  0  &  0   & 1-p_2 \\
0           & 1-r_2   &  1-r_2   & 0 
\end{pmatrix}  .
\end{align}   

 Then, the convex combination $M=\sum_i q_i M_i$ gives
     
     \begin{align} M=\begin{pmatrix}
        0&q_1r_1+q_2r_2&q_0r_0+q_2 r_2&q_0r_0+q_1r_1\\q_1p_1+q_2p_2&0&q_0 (1-r_0)+q_1p_1&q_0 (1-r_0)+q_2p_2\\q_0p_0+q_2(1-p_2)&q_0p_0+q_1(1-r_1)&0&q_1(1-r_1)+q_2(1-p_2)\\q_0(1-p_0)+q_1(1-p_1)&q_0(1-p_0)+q_2(1-r_2)&q_1(1-p_1)+q_2 (1-r_2)&0
    \end{pmatrix}.
    \end{align}
    
    Then, the simulation condition (\ref{simulatenot2}) implies the linear system of equations
     \begin{align}
        \nonumber  [M]_{12}+[M]_{21}+[M]_{34}+[M]_{43}  &=2(q_1+q_2)=\frac 43\\
        [M]_{14}+[M]_{41}+[M]_{23}+[M]_{32}&=2(q_0+q_1) =\frac 43  \, .
    \end{align}
    The only solution to this system is $q_0=q_1=q_2=\frac 13$.   \qed

\section{A simulation protocol using perfectly correlated, uniformly random trits}\label{IIIA}

Here we provide an explicit protocol that simulates   the $4$-dimensional {\tt NOT} channel  using one bit of classical communication and  two perfectly correlated, uniformly random trits. 

The protocol is as follows: 
\begin{enumerate} 
\item  Alice encodes the input variable $b\in  \{1,2,3,4\}$ into a  bit, using a a function   $f^{(i)}(b)$ that depends on the value of  her trit, denoted by $i\in  \{1,2,3\}$.  Specifically, the action of the encoding channel on the input variable is
\begin{align}\label{EtA}
f^{(i)} (b)  :  =  
\left\{    \begin{array}{ll}    
1  \qquad \qquad  &  {\rm if}~b  \in  \{1,1+ i\} \\ 
&\\
0  \qquad  \qquad &  {\rm if}~b  \not \in  \{1,1+i\}  \, .
\end{array}
\right.
\end{align}  
\item After encoding, Alice sends the message $  m : =     f^{(i)}  (b)$ to Bob using a noiseless bit channel.  
\item 
Upon receiving the message  $m$,  Bob generates the output variable $y\in  \{1,2,3,4\}$ by  applying a decoding channel   that depends on the value of his trit $j$.     
   Bob's decoding is stochastic and generates the value $y$ according to the probability distribution 
   \begin{align}
   p_{\rm dec}^{(j)}  ( y|  m)  :  =  \left\{    \begin{array}{ll}    
\frac 12   \qquad \qquad  &  {\rm if~}  m=1  ~{\rm and}~y  \not \in  \{1,1+ j\} \\
&\\
\frac 12   \qquad \qquad  &  {\rm if~}  m=0  ~{\rm and}~y \in  \{1,1+ j\} \\
&\\
0  \qquad  \qquad &  {\rm otherwise} \, .
\end{array}
\right.
   \end{align}
   \end{enumerate}
For fixed values of Alice's and Bob's trits,  the above protocol generate an output variable $y$ with probability distribution 
\begin{align}
\nonumber p^{(i,j)}(y|b)  &  =    p_{\rm dec}^{(j)} \big( y|  f^{(i)}  (b)  \big) \\
  &  =  
  \left\{    \begin{array}{ll}    
\frac 12   \qquad \qquad  &  {\rm if~}  b \in \{1,1+ i\}  ~{\rm and}~y  \not \in  \{1,1+j\} \\
&\\
\frac 12   \qquad \qquad  &  {\rm if~}  b \not \in \{1,1+i\} ~{\rm and}~y   \in  \{1,1+ j\} \\
&\\
0  \qquad  \qquad &  {\rm otherwise} \, .
\end{array}
\right.  
\end{align}
 In particular, if Alice's and Bob's trits are perfectly correlated ($i  =  j$),   the input-output distribution becomes 
 \begin{align}
\label{otherwise} q^{(i)} (y|b):  =  p^{(i,i)}(y|b)  &  = 
  \left\{    \begin{array}{ll}    
\frac 12   \qquad \qquad  &  {\rm if~}  b \in \{1,1+ i\}  ~{\rm and}~y  \not \in  \{1,1+i\} \\
&\\
\frac 12   \qquad \qquad  &  {\rm if~}  b \not \in \{1,1+ i\} ~{\rm and}~y   \in  \{1,1+i\} \\
&\\
0  \qquad  \qquad &  {\rm otherwise} \, .
\end{array}
\right.  
\end{align}

 Now, suppose that Alice's and Bob's trits are perfectly correlated and uniformly random, meaning that their joint probability distribution is 
 \begin{align}\label{uniform}
 p_{ij} =   \frac {1}3 \,   \delta_{ij} \,.
 \end{align}  
On average, the conditional distribution becomes
\begin{align}
\nonumber  p(y|b)  & : = \sum_{i,j}   \, p_{ij}  \,   p^{(i,j)}(y|b) \\
& =  \frac 13  \,   \sum_{i \in  \{1,2,3\} } \,         q^{(i)}(y|b)  \, .\label{uniformaverage}
\end{align}
To conclude, we now  show that  $p(y|b)= 1/3$ for every $y\not=  b$.    First, note that the values of $i$ are in one-to-one correspondence with the partitions of the set $\{1,2,3,4,\}$ into two-element subsets:  $i=1$ corresponds to the partition $(1,2|3,4)$,  $i=2$ corresponds to the partition $(1,3|2,4)$, and  $i=3$ corresponds to the partition $(1,4|2,3)$.   Second, note that, for every    $a\in  \{1,2,3,4\}\setminus \{b\}$,  there exists one and only one value of $i$ such that the corresponding partition contains the set $\{a,b\}$  (in other words, there exists one and only one value  $i_a$ such that either $\{a,b\}  =  \{1,1+i_a\}$ or  $\{a,b\}  =  \{1,2,3,4\} \setminus \{1,1+i_a\}$).     For this value,  Eq. (\ref{otherwise}) yields
\begin{align}\label{bastaaa}
q^{(i_a)}(y|b)   =  \frac 12   \qquad \forall y  \in  \{1,2,3,4\} \setminus \{a,b\} \, .
\end{align} 
   Let us enumerate the possible values of  $a\in  \{1,2,3,4\}\setminus \{b\}$ as $\{a_1,a_2,a_3\}$ and the corresponding values of $i_a$ as $\{i_1,i_2,i_3\}$. With this notation,  Eq. (\ref{bastaaa}) can be rewritten as  
 \begin{align}
\nonumber q^{(i_1)}(y|b)   &=  \frac 12  (\delta_{ y, a_2 }  +  \delta_{y, a_3}  )    \\
\nonumber q^{(i_2)}(y|b)   &=  \frac 12  (\delta_{ y, a_1 }  +  \delta_{y, a_3}  )    \\
 q^{(i_3)}(y|b)   &=  \frac 12  (\delta_{ y, a_1 }  +  \delta_{y, a_2}  )   \,, 
\end{align} 
where $\delta_{y,a}  $ is the Kronecker delta  ($\delta_{y,a}  =1$ if $y=a$ and $\delta_{y,a}  =0$ if $y\not  =  a$). Hence,  the probability distribution  (\ref{uniformaverage}) can be rewritten as  
\begin{align}
\nonumber  p(y|b)  & =   \frac { q^{(i_1)}(y|b)  +  q^{(i_2)}(y|b)    +  q^{(i_3)}(y|b)    }3\\  
\nonumber &=  \frac { \frac 12  (\delta_{ y, a_2 }  +  \delta_{y, a_3}  )  +    \frac 12  (\delta_{ y, a_1 }  +  \delta_{y, a_3}  )  +  \frac 12  (\delta_{ y, a_1 }  +  \delta_{y, a_2}  )    }3\\
&=  \frac {  \delta_{ y, a_1 }  +  \delta_{y, a_2}  + \delta_{ y, a_3 }     }3 \, .
\end{align}
Hence, we have shown that $p(y|b)  =1/3$ whenever $y\not  =  b$.   Combined with the normalization of the probability distribution $p(y|b)$ this condition also implies $p(y|b)=  0$ if $y =  b$.   

In summary,  we have proven than $p(y|b)$ coincides with the probability distribution $p_{\tt NOT}  (y|b)$ in Eq. (\ref{classicalNOT}), meaning that above protocol simulates the $4$-dimensional {\tt NOT} channel.  The simulation uses one bit of noiseless classical communication and two perfectly correlated trits with uniform probability distribution  (\ref{uniform}).

\section{Simulation of the universal {\tt NOT} channel}
\label{app:unot}

Here we show an explicit protocol that simulates the {\tt UNOT} channel using one bit of classical communication and $\log 3$ bits of classical shared randomness. The protocol is as follows:  

\begin{protocol}[Simulation of the universal {\tt NOT} channel]\label{protocol:unotsimulation}
Before the beginning of the protocol, Alice and Bob share two perfectly correlated trits, with uniformly distributed values.
\begin{enumerate}
\item If the value of the trit is $i \in \{1,2,3\}$,  Alice measures the input qubit in the orthonormal basis ${\sf B}_i =  \{|\psi^{(i)}_0\>, |\psi^{(i)}_1\>\}$ corresponding to the eigenstates of the Pauli matrix $\sigma_i$, with $\sigma_1: =  X$, $\sigma_2: =  Y$, and $\sigma_3: =  Z$, 
\item Alice uses the classical bit channel to communicate the outcome of her measurement to Bob,
\item Upon receiving the measurement outcome $m \in  \{0,1\}$, Bob prepares the output qubit in the basis state $|\psi^{(i)}_{m\oplus 1}\>$, depending on the outcome $m$ and on the trit value $i$.  
\end{enumerate}
\end{protocol}

On average, the above protocol  yields the {\tt UNOT} channel.  To see this, one can start from the  eigenvalue condition    
\begin{align}
 \sigma_i  \,   |\psi^{(i)}_m\> =  (-1)^m    |\psi^{(i)}_m\> \, ,  \qquad  \forall m\in  \{0,1\}, \, \forall i\in  \{1,2,3\}    \, ,
\end{align}
which implies the relations  
\begin{align}
 \nonumber  |\psi^{(i)}_m\>  \<\psi^{(i)}_m| &  =  \frac{   I  +  (-1)^m \,  \sigma_i}2  \\
|\psi^{(i)}_{m\oplus 1}\>  \<\psi^{(i)}_{m\oplus 1}| &  =  \frac{   I  -  (-1)^m \,  \sigma_i}2 \,.
\end{align}

Suppose that Alice measures and Bob prepares states  in the basis of the eigenstates of $\sigma_i$.  Then, the output state is
  \begin{align}
\nonumber \sum_{\substack{m \in  \{0,1\}}}    \bra{\psi_m^{(i)}}\rho \ket{\psi_m^{(i)}} \ketbra{\psi^{(i)}_{m\oplus 1}}{\psi^{(i)}_{m\oplus 1
    }}      &  =  \sum_{m\in  \{0,1\}}  \Tr  \left[ \frac{ I  +  (-1)^m  \sigma_i  }2  \, \rho  \right]~ \frac{ I  - (-1)^m  \sigma_i  }2\\
    \nonumber  &  =  \sum_{m\in  \{0,1\}}     \frac{ 1  +  (-1)^m   \Tr [\sigma_i\, \rho]  }2     ~  \frac{ I  - (-1)^m  \sigma_i  }2\\
      &  =  \frac{I-  \Tr  [\sigma_i \, \rho]\, \sigma_i } 2 \, ,
  \end{align}
Averaging over the three possible bases with uniform probability,  we then obtain the output state 
\begin{align}
\frac 13  \,  \nonumber \sum_{\substack{m \in  \{0,1\}}}    \bra{\psi_m^{(i)}}\rho \ket{\psi_m^{(i)}} \ketbra{\psi^{(i)}_{m\oplus 1}}{\psi^{(i)}_{m\oplus 1
    }}     &=  \frac 13 \,  \sum_i     \frac{I-  \Tr  [\sigma_i \, \rho] \, \sigma_i} 2   \\
  \nonumber   &  = \frac I 2    - \frac 13    \left( \frac{ I +   \Tr [X \rho]\, X + \Tr [  Y  \rho] \, Y   +  \Tr[Z \rho]\,  Z   }2   -\frac{  I}  2   \right) \\
   \nonumber   &  = \frac I 2    - \frac 13    \left( \rho    -\frac{  I}  2   \right)  \\
   \nonumber &  =  \frac{2}3  \,I    -  \frac 13  \rho \\
   &  =  {\tt UNOT}  (\rho) \, ,
\end{align}
where the second equality follows from the expansion  $ O  =  (\Tr  [O]\,  I +  \Tr [X\, O] \,X +  \Tr [Y  \, O] \, Y +  \Tr [Z\, O] \,  Z )/2$ of an arbitrary operator $O \in L(\C^2)$ over the   basis $\left\{ \frac I{\sqrt 2}  ,\,  \frac{X}{\sqrt 2}   ,\,  \frac{Y}{\sqrt 2}    , \, \frac{Z}{\sqrt 2} \right\}$, which is orthonormal with respect to the Hilbert-Schmidt product. 
Since the equality holds for every input state $\rho$,  the above protocol simulates the $\tt UNOT$ channel.

\section{Simulation of entanglement-breaking qubit channels}
\label{app:entbreak}

Here we show that every entanglement-breaking qubit channel can be simulated using a bit of classical communication assisted by shared randomness.

Entanglement-breaking  channels  \cite{Horodecki2003}  form a convex set.    In the qubit case, Ruskai has shown that    the extreme points of this convex set are  CQ channels   \cite{Ruskai2003}, that is, channels of the form 
\begin{eqnarray}
  \map N_{\rm CQ}(\rho) = \sigma_0 \, \langle \psi_0|\rho|\psi_0\rangle + \sigma_1\,  \langle \psi_1|\rho|\psi_1\rangle \, ,
  \label{eq:cqq}
\end{eqnarray}
where $\{|\psi_0\> ,  |\psi_1\>\}$ is an orthonormal basis  and, for every $m\in  \{0,1\}$,  $\sigma_m $ is a density matrix.    
Hence, every entanglement-breaking qubit channel $\map N_{\rm EB}$  can be written as a convex combination of  CQ  channels, namely  
\begin{eqnarray}
\map N_{\rm EB} (\rho)= \sum_{i=1}^{K}   \,q_i  \,   \map N_{\rm CQ}^{(i)}(\rho) \, ,
\label{eq:ebcqexpansion}
\end{eqnarray} 
where $K$ is a positive integer,  $(q_i)_{i=1}^K$ is a probability distribution, and  
\begin{eqnarray}
  \map N_{\rm CQ}^{(i)}(\rho) = \sigma^{(i)}_0 \, \langle \psi^{(i)}_0|\rho|\psi_0^{(i)}\rangle + \sigma_1^{(i)}\,  \langle \psi_1^{(i)}|\rho|\psi_1^{(i)}\rangle \, ,
  \label{eq:cqq}
\end{eqnarray}

We now provide a protocol that allows a sender (Alice) and a receiver (Bob)  to simulate the transmission of a quantum system through channel  $\map N_{\rm EB}$ using one bit of classical communication, and a pair of perfectly correlated $K$-dimensional classical systems.

 \begin{protocol}[Simulation of  qubit entanglement-breaking channels]
 \label{protocol:simulationQEB}
 This protocol simulates the transmission of a qubit through  channel  $\map N_{\rm EB}$. 
  Initially,  Alice and Bob share a pair of classical $K$-dimensional systems  with probability distribution $p_{AB}  (i,j)   =  q_j  \,  \delta_{ij}$, where $(q_i)_{i=1}^K$ is the probability distribution  in  Eq. \eqref{eq:ebcqexpansion}.  The simulation protocol is as follows: 
     \begin{enumerate}
         \item  Alice has a qubit in a generic state $\rho$.    If  her classical system is in the state $i$,  Alice  measures the qubit in the orthonormal basis $\{|\psi^{(i)}_0\> ,  |\psi^{(i)}_1\>  \}$, obtaining an outcome $m\in  \{0,1\}$.  
         \item Alice  communicates $m$ to Bob, using a classical bit channel.  
         \item Upon receiving message $m$,  Bob prepares the qubit state $\sigma^{(i)}_{m}$, depending on the value of the message, and on the state $i$ of his classical system. 
     \end{enumerate}
 \end{protocol}
For every $i \in  \{1,\dots,  K\}$,  the above protocol reproduces the action  of  the  CQ channel  $\map N_{CQ}^{(i)}$ in  Eq.  (\ref{eq:cqq}).  On average over the possible values of $i$,  it reproduces the convex combination in the r.h.s. of Eq. \eqref{eq:ebcqexpansion}.   

An upper bound on the dimension of the classical systems shared by Alice and Bob can be obtained from  Caratheodory's theorem \cite{cara},  which ensures that the number of extreme points appearing in the decomposition of an element of a finite-dimensional convex set is at most $D+1$, where $D$ is the dimension of the vector space in which the convex set is embedded.  For quantum systems of dimension $d$, the convex set of entanglement-breaking  channels is included in the set of all qubit channels, which has dimension $D  = d^4 -  d^2$. In particular, for $d=2$ the vector space has dimension $12$.    Hence, Caratheodory's theorem guarantees that the convex decomposition  \eqref{eq:ebcqexpansion} contains at most $13$ terms, and that the mutual information between Alice and Bob is at most $\log 13$.  

Interestingly, our results on the bomb-and-prize game imply that the noisy version of Ruskai's theorem does not hold:  an entanglement-breaking channel with entanglement-assisted capacity $<1$ cannot, in general, be decomposed as a random mixture of CQ channels with capacity $<1$.  Every quantum {\tt NOT} channel provides a counterexample:
\begin{proposition}
Let $\map N_{\rm eb}$  a quantum $\tt NOT$ channel and  $\map N_{\rm eb}  =  \sum_j\,  q_j  \map N^{(j)}_{\rm cq}$ be an arbitrary decomposition $\map N_{\rm eb}$ into a random mixture of  CQ channels.    Then, $C_{\rm E}  (\map  N_{\rm cq}^{(j)})   =  1$ for every $j$ such that $q_j>0$.   
\end{proposition}

\Proof Since $\map N_{\rm eb}$ is a quantum $\tt NOT$ channel,  Proposition \ref{lem:notimpliesavoid} guarantees that $\map N_{\rm eb}$ satisfies the bomb-avoiding condition. For every bomb-avoiding protocol using channel $\map N_{\rm eb}$, the probability of opening the box with the bomb satisfies the condition  
\begin{align}
\nonumber 0  &=  p_{\map N_{\rm eb}}  (b|b,x)  \\
  &   =  \sum_j\,  q_j\,  p_{\map N_{\rm cq}^{(j)}}  (b|b,x)  \, ,
\end{align}
which implies $ p_{\map N_{\rm cq}^{(j)}}  (b|b,x)  =  0$ for every $j$ such that $q_j  \not  =0$.    Hence, Proposition \ref{lem:almostthere} implies that each CQ channel $\map N_{\rm cq}^{(j)}$ must be unitarily equivalent to a  $\tt NOT$ channel, which is possible only if  $\map N_{\rm cq}^{(j)}  (\rho) =  \sum_{m\in  \{0,1\}}  |\phi^{(j)}_{m}\>\< \phi^{(j)}_{m}|  \,  \<\psi^{(j)}_m|  \rho  |\psi^{(j)}_m\>$ for some bases $\{|\psi^{(j)}_0\>,  \, |\psi^{(j)}_1\> \}$ and $\{|\phi^{(j)}_0\>,  \, |\phi^{(j)}_1\> \}$.  Hence, the classical (and entanglement-assisted) capacity of $\map N_{\rm cq}^{(j)}$ is equal to 1. \qed   
\end{widetext}

\end{document}